\newif\ifplainstyle
\plainstyletrue
\plainstylefalse

\newif\ifjhepstyle
\jhepstyletrue

\newif\ifprstyle
\prstyletrue
\prstylefalse

\ifprstyle
	\documentclass[twocolumn,nofootinbib]{revtex4-1}
\else
	\documentclass[11pt,a4paper]{article}
\fi

\ifjhepstyle
	\usepackage{jheppub}
	\usepackage{amsfonts}
	\usepackage{verbatim}
	\usepackage{float}
	\usepackage{color}
	\usepackage{array}
	\newcolumntype{C}[1]{>{\centering\arraybackslash$}p{#1}<{$}}
	\makeatletter
	\def\@fpheader{\phantom{Prepared for submission to JHEP}}
	\makeatother
\else	
 	\ifprstyle
		\usepackage{verbatim}
		\usepackage{amsmath,amsfonts,amssymb}
		\usepackage[colorlinks=true
                	,urlcolor=blue
                	,anchorcolor=blue
                	,citecolor=blue
                	,filecolor=blue
                	,linkcolor=blue
                	,menucolor=blue
                	]{hyperref}
	\else
            	\usepackage{verbatim}
            	\usepackage{cite}
            	\usepackage{setspace}
            	\usepackage[top=2.5cm, bottom=2.75cm, left=2.5cm, right=2.5cm]{geometry}
            	\usepackage{amsmath,amsfonts,amssymb}
            	\usepackage[colorlinks=true
            	,urlcolor=blue
            	,anchorcolor=blue
            	,citecolor=blue
            	,filecolor=blue
            	,linkcolor=blue
            	,menucolor=blue
            	,linktoc=page
            	]{hyperref}
            	\usepackage{float}
            	\restylefloat{table}
            	
            	\numberwithin{equation}{section}
	\fi
\fi

\allowdisplaybreaks

\usepackage{mathtools}
\usepackage{subcaption}
\usepackage{bbold}
\usepackage{transparent}
\usepackage[normalem]{ulem}

\newcommand{\ThisIsTheTitle}{Bootstrapping holographic warped CFTs or: how\\
I learned to stop worrying and tolerate negative norms} 
\newcommand{\ThisIsAuthorOne}{Luis Apolo}
\newcommand{\ThisIsEmailOne}{apolo@math.tsinghua.edu.cn}
\newcommand{\ThisIsAuthorTwo}{and Wei Song}
\newcommand{\ThisIsEmailTwo}{wsong@math.tsinghua.edu.cn}
\newcommand{\ThisIsTheAffiliation}{Yau Mathematical Sciences Center, Tsinghua University, Beijing 100084, China}
\newcommand{\TheseAreTheKeywords}{}

\newcommand{\ThisIsTheAbstract}{We use modular invariance to derive constraints on the spectrum of warped conformal field theories (WCFTs) --- nonrelativistic quantum field theories described by a chiral Virasoro and $U(1)$ Kac-Moody algebra. We focus on holographic WCFTs and interpret our results in the simplest holographic set up: three dimensional gravity with Comp\`ere-Song-Strominger boundary conditions. Holographic WCFTs feature a negative $U(1)$ level that is responsible for negative norm descendant states. Despite the violation of unitarity we show that the modular bootstrap is still viable provided the (Virasoro-Kac-Moody) primaries carry positive norm. In particular, we show that holographic WCFTs must feature either primary states with negative norm or states with imaginary $U(1)$ charge, the latter of which have a natural holographic interpretation. For large central charge and arbitrary level, we show that the first excited primary state in any WCFT satisfies the Hellerman bound. Moreover, when the level is positive we point out that known bounds for CFTs with internal $U(1)$ symmetries readily apply to unitary WCFTs.}

\ifjhepstyle
\title{\ThisIsTheTitle}

\author{\ThisIsAuthorOne}
\author{\ThisIsAuthorTwo}

\affiliation{\ThisIsTheAffiliation}

\emailAdd{\ThisIsEmailOne}
\emailAdd{\ThisIsEmailTwo}

\abstract{\ThisIsTheAbstract} 

\fi

\begin{document}


\ifjhepstyle
\maketitle
\flushbottom
\fi

\long\def\symfootnote[#1]#2{\begingroup%
\def\thefootnote{\fnsymbol{footnote}}\footnote[#1]{#2}\endgroup} 

\def\rednote#1{{\color{red} #1}}
\def\bluenote#1{{\color{blue} #1}}

\def\({\left (}
\def\){\right )}
\def\lb{\left [}
\def\rb{\right ]}
\def\lB{\left \{}
\def\rB{\right \}}

\def\Int#1#2{\int \textrm{d}^{#1} x \sqrt{|#2|}}
\def\Bra#1{\left\langle#1\right|} 
\def\Ket#1{\left|#1\right\rangle}
\def\BraKet#1#2{\left\langle#1|#2\right\rangle} 
\def\Vev#1{\left\langle#1\right\rangle}
\def\Vevm#1{\left\langle \Phi |#1| \Phi \right\rangle}\def\bbox{\bar{\Box}}
\def\til#1{\tilde{#1}}
\def\wtil#1{\widetilde{#1}}
\def\ph#1{\phantom{#1}}

\def\ra{\rightarrow}
\def\la{\leftarrow}
\def\lra{\leftrightarrow}
\def\p{\partial}
\def\diff{\mathrm{d}}

\def\sinh{\mathrm{sinh}}
\def\cosh{\mathrm{cosh}}
\def\tanh{\mathrm{tanh}}
\def\coth{\mathrm{coth}}
\def\sech{\mathrm{sech}}
\def\csch{\mathrm{csch}}

\def\a{\alpha}
\def\b{\beta}
\def\g{\gamma}
\def\d{\delta}
\def\e{\epsilon}
\def\ve{\varepsilon}
\def\k{\kappa}
\def\l{\lambda}
\def\n{\nabla}
\def\om{\omega}
\def\s{\sigma}
\def\t{\theta}
\def\z{\zeta}
\def\vp{\varphi}

\def\ss{\Sigma}
\def\dd{\Delta}
\def\GG{\Gamma}
\def\ll{\Lambda}
\def\tt{\Theta}

\def\A{{\cal A}}
\def\B{{\cal B}}
\def\C{{\cal C}}
\def\cE{{\cal E}}
\def\D{{\cal D}}
\def\F{{\cal F}}
\def\H{{\cal H}}
\def\I{{\cal I}}
\def\J{{\cal J}}
\def\K{{\cal K}}
\def\L{{\cal L}}
\def\N{{\cal N}}
\def\O{{\cal O}}
\def\P{{\cal P}}
\def\cS{{\cal S}}
\def\W{{\cal W}}
\def\X{{\cal X}}
\def\Z{{\cal Z}}

\def\mfa{\mathfrak{a}}
\def\mfb{\mathfrak{b}}
\def\mfc{\mathfrak{c}}
\def\mfd{\mathfrak{d}}

\def\we{\wedge}
\def\re{\textrm{Re}}

\def\tilw{\tilde{w}}
\def\tile{\tilde{e}}

\def\tilL{\tilde{L}}
\def\tilJ{\tilde{J}}

\def\zz{\bar z}
\def\xx{\bar x}
\def\xp{x^{+}}
\def\xm{x^{-}}

\def\VirU1{Vir \times U(1)}
\def\VirSL2R{\mathrm{Vir}\otimes\widehat{\mathrm{SL}}(2,\mathbb{R})}
\def\U1{U(1)}
\def\u1{U(1)}
\def\SL2R{\widehat{\mathrm{SL}}(2,\mathbb{R})}
\def\sl2r{\mathrm{SL}(2,\mathbb{R})}
\def\by{\mathrm{BY}}

\def\RR{\mathbb{R}}

\def\tr{\mathrm{Tr}}
\def\bnabla{\overline{\nabla}}

\def\sint{\int_{\ss}}
\def\dsint{\int_{\p\ss}}
\def\hint{\int_{H}}

\newcommand{\eq}[1]{\begin{align}#1\end{align}}
\newcommand{\eqst}[1]{\begin{align*}#1\end{align*}}
\newcommand{\eqsp}[1]{\begin{equation}\begin{split}#1\end{split}\end{equation}}

\newcommand{\absq}[1]{{\textstyle\sqrt{\left |#1\right |}}}



\ifprstyle
\title{\ThisIsTheTitle}

\author{\ThisIsAuthorOne}
\email{\ThisIsEmailOne}

\author{\ThisIsAuthorTwo}
\email{\ThisIsEmailTwo}

\affiliation{\ThisIsTheAffiliation}


\begin{abstract}
\ThisIsTheAbstract
\end{abstract}


\maketitle

\fi

\ifplainstyle
\begin{titlepage}
\begin{center}

\ph{.}

\vskip 4 cm

{\Large \bf \ThisIsTheTitle}

\vskip 1 cm

\renewcommand*{\thefootnote}{\fnsymbol{footnote}}

{{\ThisIsAuthorOne}\footnote{\ThisIsEmailOne} } and {\ThisIsAuthorTwo}\footnote{\ThisIsEmailTwo}}

\renewcommand*{\thefootnote}{\arabic{footnote}}

\setcounter{footnote}{0}

\vskip .75 cm

{\em \ThisIsTheAffiliation}

\end{center}

\vskip 1.25 cm

\begin{abstract}
\noindent \ThisIsTheAbstract
\end{abstract}

\end{titlepage}

\newpage

\fi

\ifplainstyle
\tableofcontents
\noindent\hrulefill
\bigskip
\fi

\section{Introduction and summary of results} \label{se:intro}

A fascinating aspect of holography is the possibility to constrain gravitational theories by exploiting the consistency of quantum field theories in lower dimensions. In this context, three dimensional gravity with a negative cosmological constant is a particularly useful playground. Indeed, any consistent quantum version of Einstein gravity with Brown-Henneaux boundary conditions~\cite{Brown:1986nw} is expected to be dual to a two dimensional conformal field theory (CFT)~\cite{Maldacena:1997re,Gubser:1998bc,Witten:1998qj}. Fundamental properties of two dimensional CFTs like unitarity and modular invariance can therefore be used to constrain the spectrum of putative theories of quantum gravity, see e.g.~\cite{Maloney:2007ud,Hellerman:2009bu,Keller:2014xba}.

Extending this approach beyond the context of the AdS/CFT correspondence is an important step towards understanding quantum gravity in the real world. A tractable but promising direction is to consider three dimensional gravity with alternative boundary conditions~\cite{Compere:2013bya,Troessaert:2013fma,Avery:2013dja}. In particular, Comp\`ere-Song-Strominger (CSS) boundary conditions~\cite{Compere:2013bya} lead to a theory of three dimensional gravity that is generically dual to a warped CFT (WCFT). The latter is a two dimensional, nonrelativistic quantum field theory invariant under the warped conformal transformation~\cite{Hofman:2011zj,Detournay:2012pc}
  \eq{
  \vp \to f(\vp), \qquad \qquad t \to t + g(\vp), \label{warpedtransformation}
  }
where $f(\vp)$ and $g(\vp)$ are two arbitrary functions of the angular coordinate $\vp \sim \vp + 2\pi$. Notably, the generators of eq.~\eqref{warpedtransformation} form a chiral Virasoro-Kac-Moody algebra where the $\u1$ symmetry is not an internal one since it is responsible for spacetime transformations.\footnote{While relatively new, several aspects of warped CFTs have been studied in the literature including their partition functions~\cite{Detournay:2012pc,Castro:2015uaa} and entanglement entropy~\cite{Castro:2015csg,Song:2016gtd,Song:2016pwx}. A few examples of warped CFTs are known such as chiral Liouville gravity~\cite{Compere:2013aya}, free Weyl fermions~\cite{Hofman:2014loa}, and free scalars~\cite{Jensen:2017tnb}.}

The study of WCFT was inspired by the search for holographic duals to extremal black holes. A wide class of extremal black holes --- including rotating black holes in four~\cite{Bardeen:1999px} and five dimensions~\cite{Dias:2007nj} --- feature an $SL(2,R)\times U(1)$ symmetry at the near horizon throat. With Dirichlet-Neumann boundary conditions, the asymptotic symmetries of the near horizon geometry are enhanced to a chiral Virasoro-Kac-Moody algebra~\cite{Compere:2009zj,Compere:2013bya}, suggesting that the dual field theory is a WCFT. Holographic dualities involving WCFTs have passed several checks including  a Cardy-like formula that reproduces the Bekenstein-Hawking entropy~\cite{Detournay:2012pc}, correlation functions that replicate greybody factors~\cite{Song:2017czq}, and characters that match one-loop determinants in the bulk~\cite{Castro:2017mfj}. 

Warped CFTs, like their unwarped cousins, must be invariant under modular transformations --- large diffeomorphisms of the theory defined on the torus. This fundamental aspect of WCFTs has dramatic consequences, as it relates the spectrum of the theory at large and small values of the angular potential and leads to the aforementioned Cardy-like formula~\cite{Detournay:2012pc}. In conventional CFTs, unitarity and modular invariance put further constraints on the spectrum beyond the asymptotic growth of states demanded by Cardy's formula~\cite{Hellerman:2009bu,Friedan:2013cba,Qualls:2013eha,Qualls:2015bta,Collier:2016cls}. This program is known as the modular bootstrap and it has been successfully applied to CFTs with internal (lower spin) symmetries in~\cite{Benjamin:2016fhe,Bae:2017kcl,Dyer:2017rul} and additional spacetime (higher spin) symmetries in~\cite{Apolo:2017xip,Afkhami-Jeddi:2017idc}.

In this paper we adapt the modular bootstrap to warped CFTs and use it to derive bounds on the spectrum of gravitational theories. We will focus on \emph{holographic} WCFTs, warped CFTs featuring a negative $\u1$ level, and interpret our results in the simplest gravitational setup compatible with their symmetries: three dimensional gravity with CSS boundary conditions. Besides modular invariance we assume
  \begin{itemize}
  \itemsep=0 em
  \item[(i)] Virasoro-Kac-Moody primaries with positive semidefinite norms, and
  \item[(ii)] a spectrum of primary states satisfying $h \ge p^2/k$ with $c >1$, \label{d}
  \end{itemize}
where $h$ denotes the conformal weight, $p$ is the $\U1$ charge, $k$ denotes the $\U1$ level, and $c$ is the central charge. The first assumption is critical for the validity or our results while the second is not strictly necessary but simplifies parts of the analysis. These assumptions can also be used to derive bounds on WCFTs with positive level and we show how known results for unitary CFTs with internal $\u1$ symmetries can be extended to unitary WCFTs. Conversely, the constraints on the spectrum of holographic WCFTs we are about to describe also imply constraints on CFTs with internal $\u1$ symmetries and negative level.

\bigskip

\noindent {\bf Holographic WCFTs} are characterized by a Virasoro-Kac-Moody algebra with a positive central charge but a negative $\u1$ level. Consequently, these theories feature negative norm descendant states that violate unitarity, one of the fundamental tenets of the modular bootstrap. Nevertheless, we will show that the negative norm states can be resummed into a Virasoro-Kac-Moody character whose contribution to the bootstrap equations is positive. This fact makes the modular bootstrap feasible in theories with mild violations of unitarity. Assuming positive norm Virasoro-Kac-Moody primaries we show that
\vspace{6pt}
\begin{center}
\noindent\parbox{13.7cm}{\emph{any WCFT with a Virasoro-Kac-Moody algebra and negative level must feature at least two states with imaginary $\u1$ charge.}}
\end{center}
\vspace{6pt}
This means that there is a sector in the WCFT for which the $\u1$ charge operator is nonhermitian and whose role in our story is better understood in the dual theory of gravity. Indeed, we will see that the bulk theory always contains a special state with imaginary $\U1$ charge, namely the AdS$_3$ vacuum. Other states with imaginary charge correspond to known solutions of Einstein equations with pathologies that are not always critical, as they include for instance conical defects that would hint at the necessity of matter fields in the bulk.

The presence of negative norm descendant states poses no obstruction to the derivation of bounds on the zero mode charges of the theory. In particular, it is possible to derive constraints on the conformal weight of the Virasoro-Kac-Moody characters provided the weight is bounded from below. This motivates us to consider an absolute minimum weight $h_0 = 0$ which is compatible with assumption (ii) above. Consequently, we show that the first excited Virasoro-Kac-Moody primary satisfies the Hellerman bound in the limit where the central charge is large,
  \eq{
  h < \frac{c}{12} + 0.479, \qquad \qquad c \gg 1. \label{introhellermanbound}
  }
Note that Virasoro primaries like the $U(1)$ current and related low weight operators are excluded from eq.~\eqref{introhellermanbound} as they correspond to $U(1)$ descendants of the vacuum.

\bigskip

\noindent{\bf Unitary WCFTs}, in contrast to holographic WCFTs, feature a positive $\u1$ level and positive norm descendant states. In this case we can derive a bound similar to eq.~\eqref{introhellermanbound} but stronger by a factor of $\O(10^{-3})$, a consequence of the positive norm descendant states that are absent in holographic WCFTs. Furthermore, when the $\U1$ level is positive, we can adapt the bounds derived for CFTs with internal $\u1$ symmetries in refs.~\cite{Benjamin:2016fhe,Dyer:2017rul}. Thus, we will show that positive level WCFTs satisfying assumptions (i) and (ii) must feature a large-$c$ spectrum such that~\cite{Benjamin:2016fhe,Dyer:2017rul}\footnote{Note that ref.~\cite{Dyer:2017rul} provides evidence for a stronger bound on the weight-to-charge ratio~\eqref{weighttocharge} scaling as $\sqrt{c}$. Interestingly,  such a bound can be proven analytically in theories with higher spin symmetries~\cite{Apolo:2017xip}.}
  \eq{
\frac{h}{|p|} &\le \sqrt{2\pi} \frac{c}{6} + \O(1), && \textrm{for at least one state,} \label{weighttocharge} \\
  h & < \frac{c}{12} + \O(1), \qquad&& \textrm{for the lightest charged state,} \\
 |p| &\le \frac{1}{\sqrt{2}}, && \textrm{for the smallest possible charge.}
  }
Finally, we will argue that the stronger bound for the lightest charged state derived in~\cite{Dyer:2017rul}, namely 
  \eq{
  h & < \frac{c}{\a} + \O(1), \,\, \qquad \qquad  \a > 16,
  }
should also hold in unitary WCFTs in the large central charge limit.

The paper is organized as follows. Section~\ref{se:css} describes the salient features of three dimensional gravity with CSS boundary conditions that are pertinent to our story. In Section~\ref{se:bootstrap} we introduce the partition function of the WCFT, derive the Virasoro-Kac-Moody characters, and write down the bootstrap equation. Section~\ref{se:bounds} contains our main results, where the constraints on the spectrum of holographic WCFTs are presented. The bounds on unitary WCFTs are derived in Section~\ref{se:morebounds}. In Section~\ref{se:bulk} we give a gravitational interpretation of the constraints on holographic WCFTs derived in Section~\ref{se:bounds}. We conclude in Section~\ref{se:discussion}. In Appendix~\ref{ap:dhh} we derive the nonlocal map between the Virasoro-Kac-Moody algebras discussed in Section~\ref{se:css} using the state-dependent asymptotic Killing vectors of ref.~\cite{Compere:2013bya}. In Appendix~\ref{ap:modular} we collect special values of the Dedekind eta function and the Eisenstein series.


\section{Three-dimensional gravity \`a la CSS} \label{se:css}

In this section we consider three-dimensional Einstein gravity with CSS boundary conditions and an asymptotic Virasoro-Kac-Moody algebra. We will focus on the nonstandard representation of the algebra and its map to the canonical form. We also comment on the tension between the space of allowed metric solutions and the integrability of charges, and point out its resolution.

We begin by recalling that in asymptotically AdS$_3$ spacetimes with Brown-Henneaux boundary conditions the boundary metric is flat and nondynamical~\cite{Brown:1986nw}. This boundary condition is necessary, although not sufficient, to recover the two copies of the Virasoro algebra that characterize the dual conformal field theory.\footnote{Additional boundary conditions on the subleading components of the metric are necessary to guarantee the existence of finite and nontrivial charges.} In contrast, CSS boundary conditions promote one of the components of the boundary metric $\g_{\mu\nu}$ to a dynamical (chiral) variable
  \eq{
  ds^2_{r\ra\infty} = r^2 ds^2_{b} = r^2 \g_{\mu\nu}dx^{\mu} dx^{\nu}, \qquad \qquad ds^2_{b} = - dx^+ \Big [ dx^- - B'(x^+) dx^+ \Big ],  \label{boundarymetric}
  }
where $x^{\pm} = t \pm \phi$ and $B'(x^+) = \p_+ B(x^+)$. This boundary condition, along with other boundary conditions on the subleading components of the metric, lead to an asymptotic symmetry group described by a left-moving Virasoro and $\u1$ Kac-Moody algebras~\cite{Compere:2013bya}. These are the symmetries of a WCFT and one finds that at the boundary the Virasoro-Kac-Moody symmetries act as  
  \eq{
  x^+ \ra f(x^+), \qquad \qquad x^- \ra x^- + g(x^+), \label{wcftsym}
  }
where $f(x^+)$ and $g(x^+)$ are two arbitrary functions.\footnote{The relationship between the lightcone coordinates $(x^{+},x^{-})$ and the WCFT coordinates $(t, \vp)$ used in Section~\ref{se:intro} will be clarified in the next section.} In particular, the second transformation shifts the value of $B(x^+)$ in eq.~\eqref{boundarymetric}, reason why the boundary metric must feature an arbitrary chiral function.

In the Fefferman-Graham gauge~\cite{Fefferman:1985ok} it is possible to write the most general solution to Einstein's equations compatible with the boundary condition~\eqref{boundarymetric}. The space of metric solutions consistent with a well-defined variational principle is parametrized by two undetermined integration parameters, a chiral function $L(x^+)$ and a constant $\bar{L}$, which together with $B(x^+)$ yield~\cite{Compere:2013bya}
  \eqsp{
  \!ds^2 & = \frac{dr^2}{r^2} - r^2 dx^ + \Big[ dx^- - B'(x^+) dx^+ \Big ] + \frac{6}{c} L(x^+) (dx^+)^2 + \frac{6}{c} \bar{L} \Big [ dx^- - B'(x^+) dx^+ \Big ]^2 \\
  & - \frac{1}{r^2} \frac{36}{c^2} L(x^+) \bar{L} \,dx^+ \Big [ dx^- - B'(x^+) dx^+ \Big ]. \label{metric}
  }
Comparison to the Ba\~nados parametrization of the classical phase space of AdS$_3$ gravity~\cite{Banados:1998gg} shows that the CSS boundary conditions \emph{dress up} the corresponding constant-$\bar{L}$ solutions with a boundary metric.\footnote{Ba\~nados' solution is given by eq.~\eqref{metric} with $B(x^+) \ra 0$ and $\bar{L} \ra \bar{L}(x^-)$.} In particular, we note that for constant $L(x^+) \ra L$ both the (dressed) AdS$_3$ vacuum with $L = \bar{L} = -c/24$, as well as (dressed) BTZ black holes with $L \ge  0$, $\bar{L} \ge 0$, are solutions of the theory (see Fig.~\ref{spectrum}).


\subsection{Asymptotic symmetry algebra}

There are two salient features of CSS boundary conditions that play an important role in our story. The first is that the asymptotic Virasoro-Kac-Moody symmetry algebra is not in the canonical form. Indeed, denoting the modes of the Virasoro and $\U1$ charges by $\L_{n}$ and $\P_{n}$, respectively, the asymptotic symmetry algebra reads~\cite{Compere:2013bya}
  \eqsp{
  [\L_n,\L_m] &= (n - m) \L_{n+m} + \frac{c}{12} n^3 \d_{n+m}, \\
  [\L_n,\P_m] &= - m \P_{n+m} + m \P_0\, \d_{n+m},\\
  [\P_n,\P_m] &= -2 n \P_0\, \d_{n+m}, \label{cssu1}
  }
where the central charge takes the usual Brown-Henneaux value of $c = 3\ell/2G$ --- with $\ll = -\ell^2$ the cosmological constant and $G$ Newton's constant --- and the $\L_n$ and $\P_n$ modes are given in terms of $L(x^+)$, $\bar{L}$, and $B(x^+)$ by
  \eq{
 \L_n &= \frac{1}{2\pi} \int d\phi \,e^{i n x^+} \Big \{ L(x^+) - \bar{L}\, [B'(x^+)]^2 \Big\}, \label{Ln} \\
 \P_n &= \frac{1}{2\pi} \int d\phi\, e^{inx^+}  \Big \{ \bar{L} +  2 \bar{L}\, B'(x^+) \Big \}. \label{Pn}
  }
Curiously, the $\u1$ level of the algebra depends on the value of $\P_0$, being negative for $\P_0 > 0$ and positive otherwise. In the dual WCFT this implies that some of the $\u1$ descendants of BTZ black holes --- among other $\P_0 > 0$ solutions --- have negative norm, while descendants of primary states with $\P_0 < 0$, like those of the vacuum, are healthy. Also note that the $\bar{L}$-dependent (second) term in eq.~\eqref{Ln} is almost the Sugawara contribution of the $\U1$ algebra, except that it is missing the appropriate powers of $\P_0$ modes.

\begin{figure}[!h]
  \centering
  \includegraphics{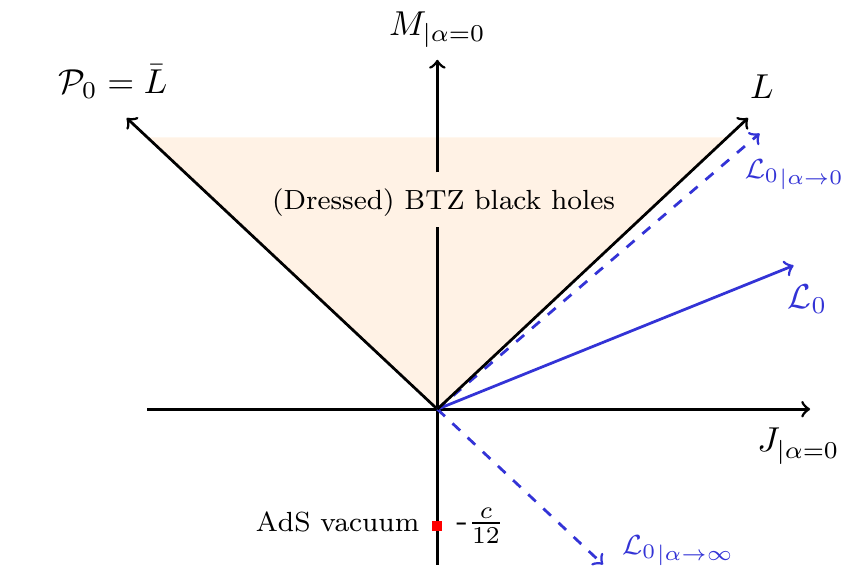}
  \caption{Spectrum of 3D gravity with CSS boundary conditions where $\L_0$ and $\P_0$ denote the zero modes of the Virasoro-Kac-Moody algebra. As $\a \ra 0$, the $\L_0$ axis approaches the $L$ axis and we recover the Brown-Henneaux identification of mass and angular momentum.}
  \label{spectrum}
  \end{figure}

As with Brown-Henneaux boundary conditions, the zero modes charges of this algebra correspond to linear combinations of mass and angular momentum, namely $\L_0  = \frac{1}{2}(M + J)$ and $\P_0  = \frac{1}{2}(M - J)$. These modes may be written as
  \eq{
  \L_0  = L - \a^2  \bar{L}, \qquad \qquad \P_0 = \bar{L}, \label{MJ}
  }
where $L$ and $\a^2$ are constants defined in terms of $L(x^+)$ and $B(x^+)$ by 
  \eq{
  L = \frac{1}{2\pi} \int d\phi L(x^+), \qquad \qquad \a^2 = \frac{1}{2\pi} \int d\phi[ B'(x^+)]^2. \label{alphaparameter}
  }
It is worth noting that if $\bar{L} \ne 0$ then $\a^2 = (2 \bar{L})^{-2} \sum_{n \ne 0} \P_n \P_{-n}$ and hence $\a^2$ is non-vanishing for backgrounds with nontrivial $\P_n$ charges. These backgrounds shift the energy and angular momentum of the corresponding undressed (Brown-Henneaux) solution as illustrated in Fig.~\ref{spectrum}. In particular, the fact that $\L_0$ rotates clockwise with respect to $L$ for increasing values of $\a^2$, i.e.~that the mass $M = \L_0 + \P_0$ decreases when $\bar{L} > 0$, is related to the fact that the level is negative whenever $\bar{L}$ is positive. On the other hand if $\bar{L} = 0$, $\a^2$ remains finite but $B(x^+)$ drops out from all charges.

Another important feature of CSS boundary conditions is that the zero mode of the $\u1$ charge is often assumed to be a fixed constant. This follows from consistency of the asymptotic symmetry group, namely the integrability of the charges~\eqref{Ln} and~\eqref{Pn}, which prevents $\bar{L}$ from varying. A fixed value of $\P_0$ seems to be too restrictive, however, as it removes many of the solutions in eq.~\eqref{metric}. Furthermore, the holographic derivation of entropy in the dual WCFT uses a Cardy-like formula that requires a chemical potential for the $\u1$ charge~\cite{Detournay:2012pc}. This implies that $\P_0$ should be allowed to vary. Relatedly, if the chemical potential is turned on, then modular invariance requires states with different $\P_0$ charges. Therefore the parameter $\bar{L}$, while still constant, should not be kept fixed --- it parametrizes different states/solutions in the same theory.

We now note that 3D gravity with CSS boundary conditions does admit integrable charges with constant but varying $\bar{L}$ provided that i) $B(x^+)$ in eqs.~\eqref{boundarymetric} and~\eqref{metric} is a periodic function; and ii) the asymptotic Killing vectors are state-dependent, i.e.~dependent on the $\P_0$ charge of the background~\cite{Compere:2013bya}. The advantage of using state-dependent Killing vectors is twofold. On the one hand they lead to a consistent asymptotic symmetry group with variable zero mode charges. On the other hand they modify the asymptotic symmetry algebra in such a way as to reproduce the canonical Virasoro-Kac-Moody algebra used in the analysis of the dual field theory (see Appendix~\ref{ap:dhh}). The bounds derived with the modular bootstrap employ this canonical Virasoro-Kac-Moody algebra. Hence, in what follows $\bar{L}$ is taken to be an arbitrary constant for which the charges of the bulk theory remain well-defined, i.e.~finite, conserved, and integrable. Note that the $\L_n$ and $\P_n$ charges given in eqs.~\eqref{Ln} and~\eqref{Pn} remain useful as their asymptotic Killing vectors are background-independent and more closely related to those of 3D gravity with Brown-Henneaux boundary conditions.


\subsection{Canonical map} \label{se:canonical}

The Virasoro-Kac-Moody algebra~\eqref{cssu1} is also featured in other holographic approaches to WCFTs~\cite{Compere:2009zj,Song:2011sr,Detournay:2012dz}, although it differs from the canonical algebra used to describe the field theory, namely
  \eqsp{
  [L_n, L_m] &= (n - m) L_{n+m} + \frac{c}{12} n^3 \d_{n+m}, \\
  [L_n, P_m] &= - m P_{n+m} ,\\
  [P_n, P_m] &=  \frac{k}{2} n \d_{n+m}. \label{canonicalu1}
  }
The two algebras are in fact related, as originally shown in ref.~\cite{Detournay:2012pc}. Indeed, after the (nonlocal) redefinition of the $\L_n$ and $\P_n$ charges,\footnote{The overall sign of $\P_n$ in eq.~\eqref{mapu1s} differs from that of ref.~\cite{Detournay:2012pc}. This is a consequence of the extra minus sign in the $\U1$ commutator of eq.~\eqref{cssu1}, which is also not present in~\cite{Detournay:2012pc}.} 
  \eq{
  \L_n = L_n + 2 P_0 P_n - P_0^2 \,\d_n, \qquad \qquad \P_n = 2 P_0 P_n - P_0^2 \,\d_n,\label{mapu1s}
  }
the CSS algebra~\eqref{cssu1} reduces to the canonical Virasoro-Kac-Moody algebra~\eqref{canonicalu1} with central charge $c = 3\ell/2G$ and level $k = -1$. In Appendix~\ref{ap:dhh} we point out an alternative way to obtain eq.~\eqref{mapu1s} using the state-dependent asymptotic Killing vectors found in~\cite{Compere:2013bya}. A crucial feature of the map~\eqref{mapu1s} is that it adds the necessary $P_n$ modes required to interpret their total contribution to $L_n$ as originating from the Sugawara construction, i.e.~we have,
  \eq{
  L_n = \frac{1}{2\pi} \int d\phi \,e^{inx^+} L(x^+) + \frac{1}{k} \sum_{m=-\infty}^{\infty} :P_m P_{n-m}:, \label{Lncanonical}
  }
where we have normal ordered the product of $P_n$ modes following the standard prescription~\cite{DiFrancesco:1997nk},
  \eq{
  \sum_{m} :P_m P_{n-m}: = \sum_{m \le -1} P_m P_{n-m} + \sum_{m > -1} P_{n-m} P_{m}. \label{normalordering}
  }
The first term in eq.~\eqref{Lncanonical} is the same contribution to the $L_n$ charge found in 3D gravity with Brown-Henneaux boundary conditions. On the other hand, the second term is consistent with the 2D gravity interpretation of ref.~\cite{Apolo:2014tua} where it was understood to be a consequence of the Weyl anomaly. In particular, if we denote the modes associated with $L(x^+)$ by $\ell_n = \frac{1}{2\pi} \int d\phi \,e^{inx^+} L(x^+)$, then the canonical Virasoro-Kac-Moody algebra implies that the $\u1$ sector of the theory factorizes, i.e.~
  \eq{
  [\ell_n , P_m] = 0. \label{lcommutator}
  }

The bounds derived in Section~\ref{se:bounds} apply to WCFTs with a canonical Virasoro-Kac-Moody algebra and are translated into constraints on the bulk theory via eq.~\eqref{mapu1s}. Before we adapt the modular bootstrap to holographic WCFTs, let us point out some interesting properties of the $L_n$ and $P_n$ modes. First, note that eq.~\eqref{mapu1s} maps $L_0$ to the bulk angular momentum, i.e.~$L_0 = \L_0 - \P_0 = J$. The WCFT bounds of Section~\ref{se:bounds} are quantum mechanical in nature and constrain the expectation value of $L_0$ on Virasoro-Kac-Moody primaries. In terms of bulk data, the expectation values of the canonical zero modes are given by
  \eq{
  \Bra{\psi} L_0\Ket{\psi} = L - \bar{L}, \qquad \qquad \Bra{\psi} P_0\Ket{\psi}  = \sqrt{\bar{L}},\label{zeromodesmap}
  }
where $\Ket{\psi}$ denotes a primary state. In eq.~\eqref{zeromodesmap} we have used the normal ordering prescription for the Sugawara modes that contribute to $L_0$~\eqref{Lncanonical}, i.e. we have used the fact that $P_n \Ket{\psi} = 0$ for $n > 0$. This ordering prescription is natural from the dual WCFT point of view. It is also consistent in the bulk theory since backgrounds with vanishing $B'(x^+)$ functions have vanishing $\P_n$ modes --- the bulk analog of $P_n \Ket{\psi} = 0$ --- and angular momentum given by $J = L - \bar{L}$, see~eqs.~\eqref{MJ} and~\eqref{alphaparameter}.

Interestingly, if all of the metrics in eq.~\eqref{metric} are interpreted as valid solutions in the quantum theory, then according to eq.~\eqref{zeromodesmap} its spectrum must include solutions with negative values of $\bar{L}$, i.e.~WCFT states with imaginary $P_0$ charge.\footnote{Note that complex values of $P_0$ lead to complex metrics in the bulk theory. Hence we will only consider real or pure imaginary $P_0$ charges, for which the metric is real.} The AdS$_3$ vacuum stands out among such states as it preserves the global $SL(2,R) \times U(1)$ symmetries of the theory. The emergence of imaginary $P_0$ charges can be traced back to the variable level of the original $\u1$ algebra~\eqref{cssu1}. Indeed, in order to fix the level of the canonical algebra to a negative constant, the map~\eqref{mapu1s} rescales some of the $\P_n$ modes by imaginary values, effectively rendering the corresponding charge an antihermitian operator. To see this note that the definition of $\P_n$ modes given in eq.~\eqref{Pn} implies that 
  \eq{
  \P_n^\dagger = \P_{-n},
  }
which follows from the assumed hermicity of the bulk $\u1$ charge. In contrast, using the map~\eqref{mapu1s}, the canonical $P_n$ modes satisfy 
  \eq{
  P_n^\dagger &= P_{-n}, \phantom{-} \qquad \qquad \textrm{if }\, P_0^2 \ge 0, \label{hermitianP}\\
  P_n^\dagger &= -P_{-n}, \qquad \qquad \textrm{if }\, P_0^2 < 0. \label{antihermitianP}
  }
As a result the canonical $\u1$ charge associated with the $P_n$ modes is hermitian for states with $\bar{L} \ge 0$ and antihermitian otherwise. From eq.~\eqref{antihermitianP} it follows that, despite the negative level, all of the $\u1$ descendants of states with $\bar{L} < 0$ have positive norm. This is a feature that carries over from the original Virasoro-Kac-Moody algebra~\eqref{cssu1} and shows that the antihermicity of the canonical $\u1$ charge for $\bar{L} < 0$ states is required for consistency. 


\section{Modular bootstrap for WCFTs} \label{se:bootstrap}

In this section we derive the modular crossing equation that is used in Section~\ref{se:bounds} to constrain the spectrum of holographic WCFTs. Unlike CFTs with internal $\U1$ symmetries there are a few subtleties we must discuss first in order to justify the applicability of our results to the bulk theory. Herein we also deal with the elephant in the room, the negative norm descendant states, and show that their overall contribution to the modular crossing equation is positive and universal, i.e.~independent of the details of the spectrum. This fact makes the modular bootstrap viable despite the violation of unitarity.


\subsection{Partition function and modular transformations}

Let us consider WCFTs with a canonical Virasoro-Kac-Moody algebra~\eqref{canonicalu1} and a negative level normalized to $k = -1$. The fact that the $\U1$ symmetry is a spacetime symmetry plays an important role in the modular properties of WCFTs. Indeed, both the $\u1$ Kac-Moody and $SL(2,R)$ symmetries are used to generate the large diffeomorphisms of the torus under which the theory must be invariant. Note that the lack of Lorentz invariance means that we have different possible tori --- parametrized by the choice of spatial cycle --- on which we can place the theory. Different choices of the spatial cycle lead to different partition functions equipped with their own modular transformation properties~\cite{Castro:2015uaa,Castro:2015csg}. However, since all of these partition functions are related to one another, we expect the constraints derived in one frame (spatial cycle) to be derivable in other frames as well. In this paper we will place the theory on the canonical circle where the spatial and thermal cycles are respectively given by
  \eq{
  (t, \,\vp) \sim (t ,\, \vp + 2\pi) ~ \sim (t + 2\pi z ,\, \vp + 2\pi\tau). \label{cycles}
  }

The coordinates ($t$, $\vp$) of the WCFT are not the same boundary lightcone coordinates ($x^{+}$, $x^{-}$) of 3D gravity with CSS boundary conditions. Neither are the two Virasoro-Kac-Moody algebras given in eqs.~\eqref{cssu1} and~\eqref{canonicalu1}. These two facts are related, as the map between the tilded and untilded generators given in eq.~\eqref{mapu1s} corresponds to a reparametrization of the coordinates~\cite{Detournay:2012pc},\footnote{This change of coordinates reproduces eq.~\eqref{mapu1s} provided $P_n \propto \d_n$, i.e.~it works only for the Virasoro and $P_0$ modes of the algebra.}
  \eq{
  x^+ = \vp, \qquad \qquad x^- = \frac{t}{2 P_0} - \vp.  \label{nonlocalmap}
  }
In particular, this change of coordinates maps the canonical circle of the WCFT into the same circle in the dual bulk theory, namely
  \eq{
  (x^+ ,\, x^-) \sim (x^+ + 2\pi ,\, x^- - 2\pi) ~ \sim (x^+ + 2\pi\g^{+} ,\, x^- + 2\pi\g^{-}),
  }
where $\g^{\pm}$ depend on $P_0$, $z$, and $\tau$. Thus, modular invariance of the bulk theory, i.e. invariance under the exchange of the spatial and thermal cycles of the torus, corresponds to modular invariance of the boundary WCFT.

In the WCFT the $L_0$ and $P_0$ modes generate translations in the $\vp$ and $t$ coordinates, respectively. Thus, the partition function is given by
  \eq{
  Z(\tau,z) = \tr(q^{L_0} y^{P_0}), \label{partitionfunction}
  }
where $q = e^{2\pi i \tau}$ and $y = e^{2\pi iz}$. Formally this is the same partition function of a chiral CFT with an internal $\U1$ symmetry. Under modular S-transformations that exchange the spatial and thermal cycles we have~\cite{Detournay:2012pc}
  \eq{
  Z(\tau', z') = e^{- \frac{i\pi}{2} \frac{z^2}{\tau} } Z(\tau, z),   \label{spartition}
  }
where $\tau'$ and $z'$ are given by
  \eq{
  \tau' = - \frac{1}{\tau}, \qquad \qquad z' = \frac{z}{\tau}. \label{stransformation}
  }
The covariance of the partition function under modular S-transformations is also shared by CFTs with internal $\U1$ symmetries~\cite{Detournay:2012pc,Kraus:2006nb}. This should not be surprising since, given the modular transformations of the potentials in eq.~\eqref{stransformation}, what ultimately determines the transformation of the partition function is the symmetry algebra~\cite{Benjamin:2016fhe}. This suggests that eq.~\eqref{spartition} does not depend on the hermicity properties of the $\u1$ charge, a fact made explicit in the geometrical derivation of eq.~\eqref{spartition} in ref.~\cite{Detournay:2012pc}. In particular, the exponential term in eq.~\eqref{spartition} is a direct consequence of the $\u1$ level, which is a fixed constant in the canonical Virasoro-Kac-Moody algebra.

Before considering the implications of modular S-transformations, note that in the WCFT modular T-transformations, $\tau \ra \tau +1$, correspond to shifts of the angular coordinate by factors of $2\pi$. Therefore, a single valued partition function must be invariant under modular T-transformations, the latter of which restrict $L_0$ to take integer values. This result is consistent with the identification of $L_0$ as angular momentum in the bulk theory which must also take integer values.


\subsection{Virasoro-Kac-Moody characters and the crossing equation}

The modular S-transformation of the partition function constrains the spectrum of the WCFT in several ways. Most dramatically, it relates the spectrum of the theory at large and small values of the angular potential $\t = -i \tau$. This leads to a version of Cardy's formula which has been successfully applied in the holographic derivation of black hole entropy in different systems generically dual to WCFTs~\cite{Detournay:2012pc}. However, as noted originally in~\cite{Cardy:1986ie,Cardy:1991kr,Hellerman:2009bu}, modular invariance --- or covariance as in our case --- of the partition function can be used to derive further constraints on the spectrum of the theory that are independent of the potentials $\tau$ and $z$.

In order to see this let us denote the eigenvalues of the $L_0$ and $P_0$ modes by $h - c/24$ and $p$, respectively. Here $h$ is the conformal weight, i.e.~the eigenvalue of $L_0$ on the plane. The partition function now reads
  \eq{
  Z(\tau,z) = \sum_{h,p} d_{h,p} \,q^{h - \frac{c}{24}} y^{p},
  }
where the sum runs over primary and descendant states of the Virasoro-Kac-Moody algebra and $d_{h,p}$ counts their norm-weighted degeneracies. Eq.~\eqref{spartition} leads to a modular crossing equation,
  \eq{
  \sum_{h,p} d_{h,p}\, \Big \{ e^{-\frac{2\pi i}{\tau} ( h - c/24 - z p )} - e^{- \frac{i\pi}{2} \frac{z^2}{\tau} + 2\pi i \tau (h - c/24 + {z} p / \tau )}\Big \} = 0, \label{naivecrossing}
  }
which must hold for all values of the potentials $\tau$ and $z$. This implies that each of the coefficients in its Taylor expansion around, say, the fixed point $\tau = i$, $z = 0$, must vanish. In other words, there is an infinite number of constraints of the form 
  \eq{
  \sum_{h,p} d_{h,p}\, (\tau\p_{\tau})^n (\p_{z})^{m} \Big \{ e^{-\frac{2\pi i}{\tau}( h - c/24 - z p )} - e^{- \frac{i\pi}{2} \frac{z^2}{\tau} + 2\pi i \tau (h - c/24 + {z} p / \tau )}\Big \} _{\tau = i, z = 0} = 0, \label{naivebootstrap}
  }
for any integers $n$ and $m$.\footnote{Note that if $m = 0$ then eq.~\eqref{naivebootstrap} yields nontrivial constraints only for odd $n$. Similarly, if $n=0$, eq.~\eqref{naivebootstrap} yields independent constraints for $m = 4 s +2$ where $s$ is a positive integer.} The modular bootstrap exploits the fact that each of the sums in eq.~\eqref{naivebootstrap} must add up to zero, and uses this information to derive bounds on the spectrum. This is achieved through linear combinations of the operators $(\tau\p_{\tau})^n (\p_{z})^{m}$ that are chosen to yield positive semidefinite summands --- which can never add up to zero --- unless certain assumptions on the spectrum of the theory are made. 

An important assumption of the modular bootstrap is unitarity, which guarantees that all the coefficients $d_{h,p}$ in eq.~\eqref{naivecrossing} are positive. This makes it possible to determine which of the linear combinations of eq.~\eqref{naivebootstrap} yields a positive-definite summand without requiring detailed knowledge of the spectrum. Negative values of $d_{h,p}$ are nevertheless admissible provided that we know the distribution of negative norm states. This is generically not possible and may go contrary to the bootstrap philosophy which aims to constrain the spectrum of CFTs with as minimal as possible a set of assumptions.

The holographic WCFTs we are interested in are not unitary and feature states with negative values of $d_{h, p}$. Indeed, a consequence of a negative level is the existence of $\U1$ descendant states with negative norm. The distribution of these negative norm states is generic and does not depend on details of the spectrum, a fact that allows us to resum all negative contributions to the partition function. In order to see this we reconsider the partition function in terms of Virasoro-Kac-Moody characters and show that the latter are manifestly positive at the fixed point $\tau = i$, $z = 0$. Thus, the corresponding crossing equation features $d_{h,p}$ coefficients whose sign depends only on the norm of Virasoro-Kac-Moody primaries.

We begin by considering the contribution of the $\U1$ descendant states to the Virasoro-Kac-Moody character.  Let $\Ket{h,p}$ denote a Virasoro-Kac-Moody primary state with conformal weight $h$ and \emph{real} charge $p$. We then note that, after an appropriate normalization, the norm of the $\U1$ descendant states satisfies
  \eq{
  \Big | \prod_{n_i, m_i} P_{-n_i}^{m_i} \Ket{h,p} \Big |^2 = \left \{ \begin{array}{ll} +1 & \textrm{if} \,\, \sum_i m_i \,\,\textrm{is even, } \\ \vspace{-8pt} \\
												-1  & \textrm{if} \,\, \sum_i m_i \,\,\textrm{is odd.} \end{array} \right. \label{norm}
  }
The distribution of negative norm states in eq.~\eqref{norm} is independent of the charge or weight of the primary state, a consequence of the abelian nature of the $\U1$ algebra. Furthermore, it is not difficult to see that all of the $\U1$ descendant states are orthogonal to each other. Thus, the contribution of the $\U1$ descendants to the Virasoro-Kac-Moody character is given by
  \eq{
  (1 - q + q^2 - \dots)(1 - q^2 + q^4 - \dots) \dotsc
   = \prod_{n=1}^{\infty} \frac{1}{1+q^n} = q^{1/24} \frac{\eta(\tau)}{\eta(2\tau)}. \label{u1descendants}
  }
In this way the negative norm states which could in principle spoil the modular bootstrap are resummed into a ratio of Dedekind eta functions $\eta(\tau)$. In order to compute the contribution of the Virasoro descendants it is convenient to use the ``Sugawara basis'' where the descendant states are obtained via 
  \eq{
  \prod_{n_i, m_i} \big(L^{(s)}_{-n_i}\big)^{m_i} \Ket{h,p},
  }
and $L_n^{(s)}$ are the modes of the Sugawara-substracted stress tensor, namely
  \eq{
   L_n^{(s)} = L_n - \frac{1}{k} \sum_{m=-\infty}^{\infty} :P_m P_{n-m}:. \label{sugawarabasis}
  }
The normal ordering prescription is given in eq.~\eqref{normalordering}. We now make the following assumption on the conformal weight and central charge,
  \eq{
  h \ge -p^2, \qquad \qquad c >1. \label{ccbound}
  }
These assumptions guarantee that the contribution of the Virasoro descendants to the Virasoro-Kac-Moody character reads
  \eq{
  \prod_{n=1}^{\infty} \frac{1}{1-q^n} (1-\d_{vac}\, q) = \frac{1}{\eta(\tau)} q^{1/24} (1-\d_{vac}\, q), 
  }
where $\d_{vac} = 1$ for the vacuum --- defined as the state annihilated by the $L_{-1}$ generator --- and is zero otherwise. 

The advantage of using the Sugawara basis is that the $L_n^{(s)}$ and $P_n$ operators commute, a fact that translates into (i) orthogonality of the corresponding descendant states, and (ii) factorization of the norm of mixed states featuring both Virasoro and $\U1$ descendants. Thus, the full Virasoro-Kac-Moody character $\chi_{h,p}(\tau,z)$ is given by the product of the Virasoro and $\U1$ contributions, whereupon\footnote{We thank Diego Hofman for discussions on this character.}
  \eq{
  \chi_{h, p} (\tau,z) = \frac{1}{\eta(2\tau)} \, q^{h - (c-2)/24} \, y^p (1-\d_{vac}\, q), \qquad \qquad p \in \mathbb{R}. \label{realcharacter}
  }
This character is independent of the basis used for the Virasoro descendants, but establishing eq.~\eqref{realcharacter} using the $L_n$ generators is a more elaborate task. Expanding $\chi_{h, p} (\tau,z)$ for small but positive imaginary values of $\tau$ we find
  \eq{
  \chi_{h, p} (\tau,z) = q^{h-c/24} \, y^p (1-\d_{vac}\, q) (1 + q^2 + 2 q^4 + 3 q^6 + 5 q^8 + 7 q^{10} + \O(q^{12}) ).
  }
For states other than the vacuum, the coefficients in this expansion are guaranteed to be positive semidefinite as $q^{1/12}/\eta(2\tau)$ is the generating function for the number of partitions of an integer into even parts. If the vacuum carries real (or vanishing) charge, then its character is given by eq.~\eqref{realcharacter}. Although odd powers of $q$ in the expansion of this character have negative coefficients, the character is still positive when evaluated at the fixed point $\tau = i$, $z=0$, i.e.~we have
  \eq{
  \chi_{vac,p}(i,0) \approx 0.998 \, e^{-2\pi (h - c/24)},
  }
where we have used the value of $\eta(2i)$ given in Appendix~\ref{ap:modular}. The positivity of the character at the fixed point is to be expected as descendant states make only subleading contributions proportional to powers of $e^{-2\pi}$.

In contrast to the states with real charge, descendant states with imaginary charge make only positive contributions to the corresponding Virasoro-Kac-Moody character.\footnote{From this analysis we do not find any compelling reason to include arbitrary complex charges. On the other hand, it is only for real or imaginary charges that we know how to compute the norm of descendant states. Furthermore, it is only real or imaginary charges that correspond to real metrics in the bulk theory.} Indeed, due to the antihermicity of the $P_n$ generators~\eqref{antihermitianP}, the algebra of states with imaginary charge is equivalent to a Virasoro-Kac-Moody algebra with a positive level. Hence all of the $\U1$ descendant states carry positive norm and the character is given by
  \eq{
 \chi_{h, p} (\tau,z) = \frac{1}{\eta(\tau)^2} q^{h - (c-2)/24}y^p (1 - \d_{vac} q), \qquad \qquad p \in i \mathbb{R}, \label{imaginarycharacter}
  }
where we have assumed eq.~\eqref{ccbound} still holds. Note that the factor of $(1-\d_{vac}\, q)$ in eq.~\eqref{imaginarycharacter} accounts for the fact that the vacuum may carry imaginary charge. 

Incidentally, note that it is possible to relax either or both of the assumptions made on the conformal weight and the central charge in eq.~\eqref{ccbound}. In this case, for purely imaginary values of the modular parameter $\tau = i \t$ and $\t > 0$, the character of a Virasoro-Kac-Moody primary is bounded by\footnote{It is not difficult to show that eq.~\eqref{gammabound} also implies a bound on the derivatives of the character, the latter of which would be necessary to derive bounds on the conformal weight.}
  \eq{
  \Bigg [2 -  \frac{e^{-\pi\t/12}}{\eta(i\t)} \Bigg ] \frac{e^{-\pi\t/12} \eta(i\t)}{\eta(2 i \t)} \le  e^{2\pi \t (h-c/24)} y^{-p} \chi_{h,p}(i\t,z) \le \frac{e^{-\pi\t/6}}{\eta(i\t)^2}.\label{gammabound}
  }
This bound holds for states with either real or imaginary charge. The upper bound corresponds to the Virasoro-Kac-Moody character where all Virasoro \emph{and} $\U1$ descendants have positive norm. On the other hand, the lower bound corresponds to a character where the Virasoro descendants carry negative norm, hence the factor of $2 -  \frac{e^{-\pi\t/12}}{\eta(i\t)}$, while the norm of $\U1$ descendants alternates as in eq.~\eqref{u1descendants}. In particular, the character is positive at the fixed point and satisfies
  \eq{
   0.996262  \le e^{2\pi (h - c/24)} \chi_{h,p}(i,0) \le 1.00375. \label{gammabound2}
  }
Thus, the modular bootstrap is feasible even if we relax our assumptions on the conformal weight and central charge. In particular, note that the results of Section~\ref{se:u1bounds} are valid as long as the characters are positive at the fixed point. In contrast, the Hellerman bound derived in Section~\ref{se:virasorobound} is sensitive to the values of the characters and their derivatives. There we find that different characters yield the same Hellerman bound at leading order in $c$, giving only subleading corrections of $\O(10^{-3})$.
  
To summarize, the characters of the Virasoro-Kac-Moody representations to be used in the following sections depend on the $\U1$ charge. If the charge is real, the character receives contributions from negative norm descendant states and is given by eq.~\eqref{realcharacter}. On the other hand, if the charge is imaginary all descendant states carry positive norm and the character is given by eq.~\eqref{imaginarycharacter} instead. When evaluated at the fixed point $\tau = i$, $z = 0$, the difference between these characters is of $\O(10^{-3})$.
  
In terms of Virasoro-Kac-Moody characters the partition function becomes
  \eq{
  Z(\tau, z) = \sum_{h, p} D_{h,p}\, \chi_{h,p}(\tau,z), \label{partitioncharacters}
  }
where the sum runs only over Virasoro-Kac-Moody primaries and $D_{h,p}$ accounts for their degeneracy, the latter of which is assumed to be a positive number. In particular, note that the $\U1$ current and related Virasoro primary states are already accounted for in the vacuum character since these states are $\U1$ descendants of the vacuum. Using eq.~\eqref{spartition} the modular crossing equation now reads
  \eq{
  \sum_{h,p} D_{h,p} \,F(h,p) \equiv \sum_{h,p} D_{h,p} \Big \{ \chi_{h, p} (-1/\tau,z/\tau) - e^{- \frac{i\pi}{2} \frac{z^2}{\tau} }\chi_{h, p} (\tau, z) \Big \} = 0.\label{crossing0}
  } 
This form of the crossing equation is serviceable when derivatives with respect to $\tau$ are taken. Since these derivatives act nontrivially on the characters and the latter differ for states with real or imaginary charges, further manipulations of the crossing equation are not particularly helpful when $\tau$ derivatives are involved. On the other hand, when only derivatives with respect to $z$ are taken, a more useful formulation of the crossing equation is given by
  \eq{
 \sum_{h, p} D_{h,p} \g_{h,p}(-1/\tau) \, \tilde{F}(h,p) = 0, \label{crossing}
  }
where $\tilde{F}(h,p)$ reads
  \eq{
  \tilde{F}(h, p) = \tilde{\chi}_{h, p} (-1/\tau,z/\tau) - \frac{\g_{h,p}(\tau)}{\g_{h,p}(-1/\tau)} e^{- \frac{i\pi}{2} \frac{z^2}{\tau} }  \tilde{\chi}_{h, p} (\tau, z), \label{Fdefinition}
  }
and we have defined
  \eq{
  \tilde{\chi}_{h,p}(\tau,z) = q^{h} y^{p}, \qquad \qquad \g_{h,p} (\tau) = \frac{\chi_{h,p}(\tau,z)}{\tilde{\chi}_{h,p}(\tau,z)}. \label{gammadefinition}
  }
The main advantage of this parametrization, beyond the simplification of future equations, is that it makes manifest the fact that detailed knowledge of the $\g_{h,p}(\tau)$ function is not necessary when derivatives with respect to $z$ are taken. Indeed, the prefactors accompanying $\tilde{F}(h,p)$ in eq.~\eqref{crossing} are always positive at $\tau = i$. Likewise, at the fixed point the ratio of $\g_{h,p}(\tau)$ functions in eq.~\eqref{Fdefinition} equals 1. We conclude that the existence of negative norm states does not impose an obstruction to the modular bootstrap as their contribution to the crossing equation is generic and can be resummed into a positive number.


\section{Holographic WCFT bounds} \label{se:bounds}  

Let us now use modular covariance of the partition function to derive constraints on the spectrum of holographic WCFTs and, correspondingly, 3D gravity with CSS boundary conditions. The modular crossing equation imposes an infinite number of constraints on the spectrum of the theory, which follow from the fact that eq.~\eqref{crossing} must hold for all values of the potentials $\tau$ and $z$. We write these constraints as
  \eq{
  \sum_{h, p} D_{h, p}\, {\dd} F(h, p) = 0, \label{bootstrap}
  }
where $\dd$ is a differential operator evaluated at $\tau = i$, $z = 0$ that is given by
  \eq{
  \dd = \sum_{n,m} b_{n,m}  (\tau \p_{\tau} )^{n}  \p_{z}^{\,m}, \label{Loperator}
  }
with real $b_{n,m}$ coefficients. We restrict our analysis to operators $\dd$ with a small total number $N \le 6$ of derivatives and hence consider only the first few constraints imposed by the modular crossing equation~\eqref{crossing}. Nevertheless, we will see that these constraints are powerful enough to impose nontrivial bounds on the spectrum.


\subsection{States with imaginary \texorpdfstring{$\u1$}{U(1)} charge} \label{se:u1bounds}

Although holographic WCFTs feature a negative $\u1$ level, and are therefore not unitary, we may still hope to restrict the $\u1$ charges to be real valued. This is inconsistent with modular invariance which demands the existence of states with imaginary charge if the $D_{h,p}$ coefficients are positive. In order to see this let us consider the tilded representation of the crossing equation~\eqref{crossing} --- to be used exclusively in this subsection --- and one of the simplest differential operators, $\dd_1 = \p_z^{\,2}$, which must satisfy
  \eq{
 0 & = \sum_{h, p} D_{h, p} \g_{h,p}(i) \dd_1\, \tilde{F}(h, p) ,\\
 0 & = \sum_{h, p} D_{h, p}\g_{h,p}(i) e^{-2 \pi h} \pi (1 + 8 \pi  p^2 ). \label{delta1}
  }
Since the summand is positive definite for states with real charge we reach a contradiction and find that 
\vspace{5pt}
 
 \begin{itemize}
 \item[\emph{(i)}]\emph{at least one Virasoro-Kac-Moody primary with either negative norm or imaginary charge $p^2 < -1/8\pi$ must be included in the spectrum.}
 \end{itemize}

\vspace{5pt}
\noindent In the context of CFTs with an internal $\U1$ symmetry this result is consistent with the expectation of negative norm primary states. However, holographic WCFTs suggest the second option is also viable and, in the context of the modular bootstrap, the more interesting alternative. Indeed, as discussed in Section~\ref{se:css} the bulk theory of gravity features an exceptional state with imaginary $\u1$ charge --- the AdS$_3$ vacuum --- which is invariant under the $SL(2,R) \times U(1)$ global symmetries of the theory and minimizes the combination of zero modes $h + p^2$. This motivates us to consider the existence of at least one state with imaginary $\u1$ charge in the spectrum. We will parametrize the charge of this state by
  \eq{
  p_{a}^{\,2} = -d_a, \qquad \qquad d_a > 0,
  }
while the corresponding conformal weight $h_a$ is left arbitrary. Since the contribution of this state to the crossing equation is a special one, it will prove convenient to distinguish it from the contribution of the primary states with real charge. Thus, the crossing equation may be written as
  \eq{
  \sum_{h, p \ne h_a, p_a} \tilde{D}_{h, p} \tilde{\g}_{h,p}(-1/\tau) \, \dd  \tilde{F}(h, p) = - \dd 
  \tilde{F}(h_a, p_a), \label{crossing2}
  }
where $\dd$ is an operator featuring only derivatives with respect to $z$ and we have defined for convenience 
  \eq{
   \tilde{D}_{h, p} = \frac{D_{h, p}}{D_{h_a, p_a}}, \qquad \qquad \tilde{\g}_{h,p}(-1/\tau)  = \frac{\g_{h,p}(-1/\tau)}{\g_{h_a,p_a}(-1/\tau)}.
  }  
  %
  

\bigskip

\subsubsection*{More states with imaginary \texorpdfstring{$\u1$}{U(1)} charge}

\begin{figure}[!h]
    \centering
    \begin{subfigure}{0.5\textwidth}
        \centering
        \includegraphics{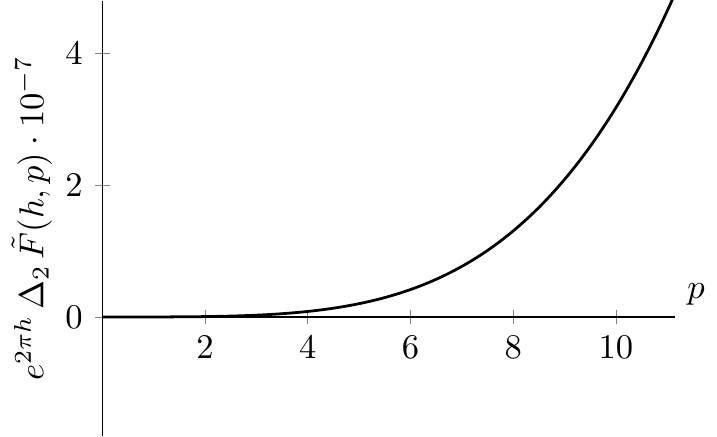}
        \caption{}
        \label{genericp06}
    \end{subfigure}%
    \begin{subfigure}{0.5\textwidth}
        \centering
        \includegraphics{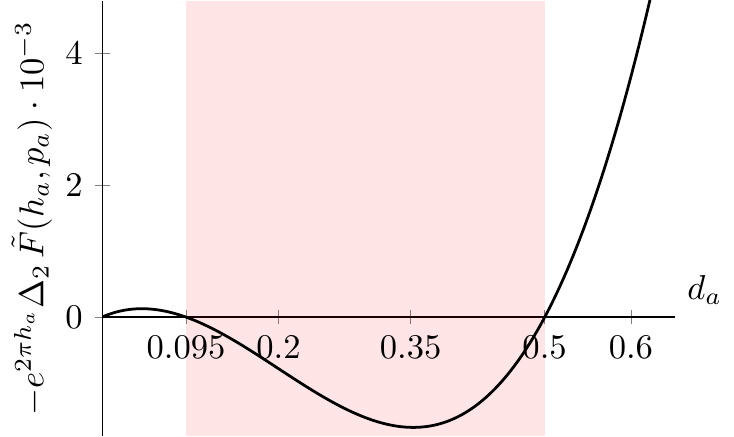}
        \caption{}
        \label{vacuump06}
        \end{subfigure}
    \caption{The contribution to the crossing equation~\eqref{crossing2} from (a) states with real-valued charges and (b) the state with imaginary charge $p_a{}^2 = -d_a$. If $d_a$ lies in the shaded region, the WCFT must contain an additional state with imaginary charge.}
    \label{p06plot}
  \end{figure}

The $(h_a,p_a)$ state described above cannot be the only state with imaginary $\u1$ charge. Indeed, modular invariance of the WCFT requires at least one additional operator with a negative value of $p^2$. In order to show this we need to consider operators with a greater number of $\p_z$ derivatives and no derivatives in $\tau$. We first note that not all of the $\p_z^{\,n}$ operators yield linearly independent $\p_z^{\,n} \tilde{F}(h,p)$ functions. The set of linearly independent real functions $\p_z^{\,n} \tilde{F}(h,p)$ is generated by
  \eq{
  \tilde{\dd}^{(s)} = \p_z^{\,4 s + 2}.
  }
It is not difficult to show that $\tilde{\dd}^{(s)} \tilde{F}(h,p)$ is a positive semidefinite polynomial in $p^2$, i.e.~we have
  \eq{
   \tilde{\dd}^{(s)} \tilde{F}(h,p) = \sum_{n=1}^{2s+1} a_n \,p^{2n}, \qquad \qquad a_{2s+1} > 0, \quad a_{i\ne2s+1} \ge 0.
  }
In particular, this implies that for operators $\dd = \tilde{\dd}^{(s)}$ the summand on the left hand side (LHS) of the crossing equation~\eqref{crossing2} is positive for states with real charge. On the other hand, for large values of $d_a$, the contribution of the $(h_a, p_a)$ state to the crossing equation is manifestly negative,
  \eq{
  \til{\dd}^{(s)}  \tilde{F}(h_a,p_a) \sim -d_{a}{}^{2s+1}, \qquad \qquad d_a \gg 1.
  }
This means that the crossing equation, as written in eq.~\eqref{crossing2}, can be easily satisfied for large enough values of $d_a$ without any additional assumptions on the spectrum. However, since $-\til{\dd}^{(s)} \tilde{F}(h_a,p_a)$ is not positive everywhere, there are ranges $D^{(s)}_a$ in the space of $d_a$ charges where the LHS of eq.~\eqref{crossing2} is positive while the right hand side (RHS) is negative (see Fig.~\ref{p06plot}). Hence, for values of $d_a \in \cup_n D_a^{(n)}$ the crossing equation becomes inconsistent unless an additional primary state with imaginary charge is added to the spectrum.

We can illustrate this with a simple example using the operator $\dd_2$ defined by
  \eq{
  \dd_2 = \p_z^{\,6} - 15 \pi^2 \p_z^{\,2}, \label{p6}
  }
where the relative coefficient is chosen to maximize the bounds we are about to describe. The action of this operator on $\tilde{F}(h,p)$ is given by
   \eq{
   \dd_2  \, \tilde{F}(h,p) &= e^{-2 \pi h} (4 \pi^4 p^2)(15 +60 \pi  p^2 + 32 \pi^2 p^4), \label{delta2}
   } 
which is not positive definite for states with arbitrary imaginary charge. Indeed, as illustrated in Fig.~\ref{p06plot}, $- \dd_2 \, \tilde{F}(h_a,p_a)$ is negative whenever $d_a \in D^{(2)}_a$ where
  \eq{
  D^{(2)}_a \approx (0.095, 0.502). \label{deltac1}
  }
Thus, modular invariant WCFTs featuring a state with $-p_a{}^2 = d_a \in D^{(2)}_a$ must contain an additional state with imaginary charge. Furthermore, it follows from eq.~\eqref{delta2} that the charge of this state must lie within
  \eq{
  -p^2 \in (0, 0.095)\cup (0.502,\infty), \label{psquaredbound}
  }
which is the complement of $D^{(2)}_a$, i.e.~the complement of the shaded region in Fig.~\ref{vacuump06}. The small range of $D^{(2)}_a$ poses an obvious obstruction to generalizing this result to states $(h_a,p_a)$ with arbitrarily large charge. In particular, note that the coefficient accompanying the $\p_z^{\,2}$ operator in eq.~\eqref{p6} is chosen to maximize $D^{(2)}_a$ so this is the best we can do at this order.

\begin{figure}[h!]
    \centering
    \begin{subfigure}{0.5\textwidth}
        \centering
        \includegraphics{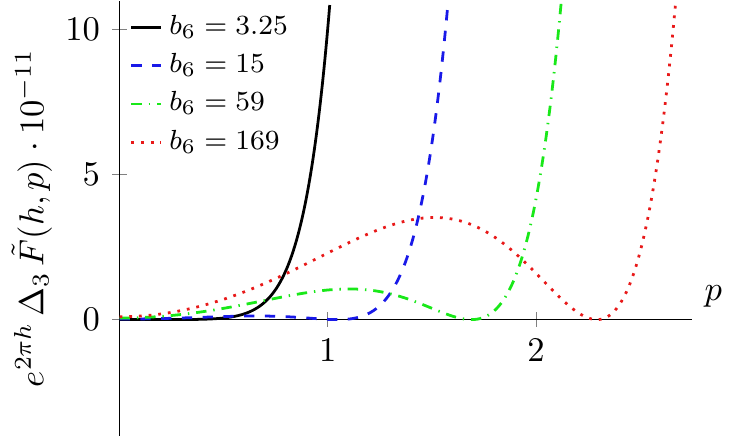}
        \caption{}
        \label{genericp10}
    \end{subfigure}%
    \begin{subfigure}{0.5\textwidth}
        \centering
        \includegraphics{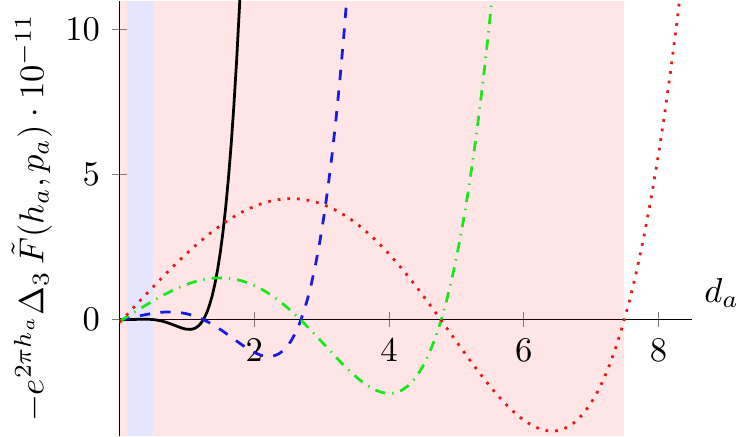}
        \caption{}
        \label{vacuump10}
        \end{subfigure}
    \caption{The contribution to the crossing equation~\eqref{crossing2} from (a) states with real-valued charges and (b) the state with imaginary charge $p_a{}^2 = -d_a$. Larger values of $b_6$ extend $D_a^{(3,i)}$ further to the right. The blue shaded region denotes the contribution of $D_a^{(2)}$ (see Fig.~\ref{vacuump06}).}
    \label{p10plot}
  \end{figure}

Let us now extend the range of $d_a$ where holographic WCFTs must feature an additional state with imaginary charge. Consider the following operator depending on two real and positive parameters $b_6$ and $b_2$,
  \eq{
  \dd_3 = \p_z^{\,10} - 63 \pi^2 b_6 \, \p_z^{\,6} + 945 \pi^4 b_2\, \p_z^{\,2}. \label{p10}
  }
For a given $b_6$, the corresponding $\dd_3\, \tilde{F}(h, p)$ function is not positive except for large enough values of $b_2$. Our strategy is to fix $b_6$ and use a semidefinite program solver~\cite{Simmons-Duffin:2015qma} to find the minimum value of $b_2$ for which $\dd_3\, \tilde{F}(h, p)$ is positive semidefinite if $p$ is real. In contrast, the contribution of the $(h_a,p_a)$ state to the RHS of eq.~\eqref{crossing2} is always negative if $d_a \in D^{(3, i)}_a$ for some range $D^{(3, i)}_a$. The $i$ index in $D^{(3, i)}_a$ is a label denoting different choices of the $b_6$ and $b_2$ parameters. While there is no value of $b_6$ for which this range is infinite, larger values of $b_6$ shift $D_a^{(3,i)}$ towards larger values (see Fig.~\ref{p10plot}). Hence, through appropriate choices of $b_6$ we can extend the region where the crossing equation becomes inconsistent to arbitrarily large values of the $d_a$ charge. We demonstrate this in Table~\ref{p10table} where $b_6$ and $b_2$ parameters are given that extend this region up to 
  \eq{
  D_a^{(2)} \cup_{i} D_a^{(3,i)} \supset (0, 7.492). \label{deltac2}
   }
  \begin{table}[h!]
  \begin{center}
  \begin{tabular}{c|c|c|rcl}
  $i$ & $b_6$ & $b_2$ & \multicolumn{3}{c}{$D_a^{(3,i)}$} \\ \hline 
  1 & 3.25 & 2.29  & $  (0, 0.106 )$ & $\cup$ & $(0.500, 1.243)$  \\
  2 & 15 & 122.58 & $ (0,  0.040 ) $ & $\cup$ & $ (1.243 , 2.692)$  \\
  3 & 59 & 2827.05 & $ (0, 0.040 )$ & $\cup$ & $ (2.684 , 4.774)$ \\
  4 & 169 & 26222 & $ (0,   0.039 )$ & $\cup$ & $ (4.766 , 7.492)$  \\
    \vdots & \vdots &\vdots & &\vdots\\
  \vdots & 10000 & 103471503 & $ (0,  0.039 ) $ & $\cup$ &  $(41.243 , 48.789) $\\
      \vdots & \vdots & \vdots& &\vdots
  \end{tabular}
  \caption{$b_6$ and $b_2$ parameters and the range $D_a^{(3,i)}$ where $-\dd_3\, \tilde{F}(h_a, p_a)$ is negative.}
  \label{p10table}
  \end{center}
  \end{table}

There is no obstruction to widening the gap~\eqref{deltac2}, as illustrated in Fig.~\ref{p10plot} and the last entry of Table~\ref{p10table}. We conclude that in holographic WCFTs with positive $D_{h,p}$ coefficients 
\vspace{4pt}
 \begin{itemize}
 \item[\emph{(ii)}]\emph{the spectrum must feature at least one additional state with imaginary $\U1$ charge.}
 \end{itemize}
 \vspace{4pt}
As with the simpler example illustrated in Fig.~\ref{p06plot}, the value of $-p^2$ for this state must lie in the compliment of $ \cup _j D_a^{(3,j)}$ where the $D_a^{(3,j)}$ must be chosen so that $d_a \in D_a^{(3, j)}$. This guarantees the existence of $b_6$ and $b_2$ coefficients for which the crossing equation breaks down. For example, for $d_a \sim 2.69$ inspection of Table~\ref{p10table} shows that $\cup _j D^{(3, j)}_a = D^{(3, 2)}_a \cup D^{(3, 3)}_a$. Hence, if $d_a \sim 2.69$ there is an additional state with imaginary charge such that
  \eq{
  -p^2 \in (0.040,1.243) \cup (4.774,\infty). \label{psquaredbound2}
  }
If the charge does note lie within this range then there exists a combination of $b_6$ and $b_2$ parameters for which the crossing equation remains inconsistent. In particular, note that the charge cannot sit too close to that of the $(h_a, p_a)$ state.

We may now parametrize the charge of this new state by $p_b{}^2 = - d_b$. The contribution of this state to the crossing equation depends on the value of $d_b$ so it is convenient to move it to the RHS of eq.~\eqref{crossing2} and attempt to repeat the analysis performed above for the appropriate $b_6$ and $b_2$ coefficients. Such a feat is generically not possible unless further assumptions on the spectrum are made. The reason is that for any coefficients $b_6$ and $b_2$ for which the contribution of $-\dd_3\, \tilde{F}(h_b, p_b)$ is negative, the corresponding contribution of $-\dd_3\, \tilde{F}(h_a, p_a)$ is positive. In general, we can not determine if the overall contribution of these states to the RHS of the crossing equation is negative. If this were the case, then modular invariance would require the addition of yet another state with imaginary charge to the spectrum --- the third of its kind. On the other hand, an overall positive contribution would constrain only the spectrum of real charges. It is possible to prove the necessity of a third state with imaginary charge for certain values of $d_a$ and $d_b$ but we cannot show this is true in general. Nevertheless, from the general properties of the functions $\til{\dd}^{(s)} \tilde{F}(h,p)$ noted above it may be possible to construct higher order operators that lead to inconsistent crossing equations unless more states with imaginary charge are added to the spectrum. 
 

\subsection{Bounds on the conformal weight} \label{se:virasorobound}

Let us now consider bounds on the conformal weight of holographic WCFTs. First, recall that the unitarity bound of a CFT with a Virasoro-Kac-Moody algebra and level $k$ is given by $h \ge {p^2}/{k}$ and $c > 1$. For $k = -1$ this is the bound given in eq.~\eqref{ccbound} that was used to derive the Virasoro-Kac-Moody characters. When the $\U1$ charge is imaginary, this bound leads to a spectrum of conformal weights that is bounded from below by $h \ra 0$. In contrast, when the charge is real, eq.~\eqref{ccbound} implies an unbounded spectrum, reason why we will further assume
  \eq{
  h \ge \left \{ \begin{array}{lccl} - p^2 &&& \hspace{15pt} \textrm{if } p^2 < 0,  \\ \vspace{-8pt} \\
\phantom{-}0 &&& \hspace{15pt} \textrm{if } p^2 \ge 0,
 \end{array} \right. \qquad\qquad c > 1.\label{unitaritybound}
  }
In particular, note that the bound for states with real charge is stronger than eq.~\eqref{ccbound} and hence it remains compatible with the derivation of the Virasoro-Kac-Moody characters of Section~\ref{se:bootstrap}. Finally, we point out that it is possible to relax the boundedness on the conformal weight and derive a bound on the spectral flow invariant combination $h - p^2/k$ instead, as discussed further in Section~\ref{se:bulk}. 

We now show that the first excited Virasoro-Kac-Moody primary in a holographic WCFT compatible with eq.~\eqref{unitaritybound} and featuring only positive norm primary states satisfies the Hellerman bound~\cite{Hellerman:2009bu}
  \eq{
  h_1 < \frac{c}{12} - \O(1), \qquad \qquad c \gg 1. \label{hellerman}
  }
Here $h_1$ denotes the weight of the first excited primary state such that $h_0 < h_1$ where $h_{0} = 0$ saturates the bound~\eqref{unitaritybound}. Note that in the dual theory of gravity with CSS boundary conditions this state does not correspond to the AdS$_3$ vacuum. Indeed, while the latter \emph{is} annihilated by the $L_{\pm 1}$ generators of the global $SL(2,R)$ transformations, it satisfies $h = c/24$ instead. In a more general context, $h_0$ would be interpreted as the vacuum, but in what follows we remain agnostic about the precise nature of this state.

Let us consider an operator $\dd_i$ acting on the untilded representation of the modular crossing equation~\eqref{crossing0}. In this equation it is convenient to split the contribution of a generic state from the state that saturates the bound~\eqref{unitaritybound}, whereby
  \eq{
  \sum_{\mathclap{h,p \ne h_0,p_0}} \, D_{h,p} \dd_i F(h,p) = - D_{h_{0}, p_0} \dd_i F(h_{0}, p_0). \label{crossing3}
  }
Restricting our analysis to the lowest number of derivatives we define the following differential operators
  \eq{
  \dd_1 = \tau \p_{\tau}, \qquad \qquad \dd_3 = (\tau \p_{\tau})^3.
  }
Then, following~\cite{Hellerman:2009bu}, the modular crossing equation~\eqref{crossing3} implies
  \eq{
  \sum_{\mathclap{h,p \ne h_0,p_0}} \, D_{h,p}\Big \{ \dd_1 F(h,p) \dd_3 F(h_{0}, p_0) -   \dd_3 F(h,p) \dd_1 F(h_{0}, p_0)\Big \} = 0. \label{crossing4}
  }
The above is a powerful equation relating several unknowns, for neither the spectrum of weights or charges, or even the characters of the representations contributing to eq.~\eqref{crossing4} need to be known. Indeed, the bound on the characters given in eq.~\eqref{gammabound} is sufficient to derive the Hellerman bound as it introduces at most an error of $\O(10^{-3})$ that is subleading in the large central charge limit. Nevertheless, in what follows we will use the precise form of the characters derived in eqs.~\eqref{realcharacter} and~\eqref{imaginarycharacter}.

Using the definition of $F(h,p)$ given in eq.~\eqref{crossing0} the modular crossing equation becomes
  \eq{
  \sum_{\mathclap{h,p \ne h_0,p_0}} \, D_{h,p} \chi_{h,p}(i) \chi_{h_0,p_0}(i)  32\pi^3  G(h,\k) = 0, \label{crossing5}
  }
where we introduced $\k = c/24$ for convenience and the factors multiplying $G(h,\k)$ are all positive semidefinite, in contrast to $G(h,\k)$ which is given by
  \eqsp
  {
  G(h,\k) =& 2\pi \k\, h^3 - 3 \k \,h^2 -   ( 3 + 2 \pi \k)\k^2 h + \dots \label{G}
  }
In eq.~\eqref{G} the dots denote terms that depend on derivatives of the characters which are subleading in the large central charge limit. These terms are approximately given by
  \eq{
  \frac{(-i)^n \p_{\tau}^n \hat{\g}_{h,p}(\tau,0)}{\hat{\g}_{h,p
  }(\tau,0)} \bigg |_{\tau = i} \sim O(10^{n-3}), \qquad \qquad n = 1,2,3,
  }
where $\hat{\g}_{h,p}(\tau,z) = q^{-(h - c/24)} y^{-p} \chi_{h,p}(\tau,z)$. Although states with real or imaginary charges are described by different characters, all characters make similar subleading contributions to the crossing equation. Nevertheless, in the calculations that follow we use the full but cumbersome expression for the $G(h,\k)$ function.
   \begin{figure}[!h]
   \begin{center}
   \includegraphics{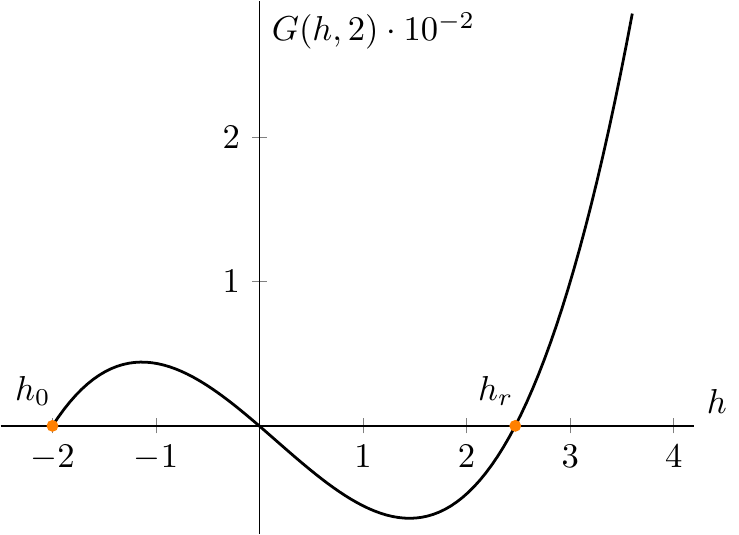}
   \end{center}
   \caption{The $G(h,2)$ function ($c = 48$) where both the $\chi_{h_0,p_0}(\tau,z)$ and $\chi_{h,p}(\tau,z)$ characters are given by eq.~\eqref{realcharacter}. The crossing equation is inconsistent unless the weight $h_1$ of the first excited state satisfies $h_1 < h_r$.}
   \label{gfunction}
   \end{figure}

The behavior of $G(h,\k)$ is illustrated in Fig.~\ref{gfunction} and features both positive and negative values, a necessary condition for the crossing equation to be satisfied. In particular, $G(h,\k)$ is positive for large enough values of $h$ and features a real zero at $h = h_{r}$ if $c > 0.838$. This restriction on $c$ is satisfied by our assumption on the central charge~\eqref{unitaritybound}. Hence, if we assume that the first excited Virasoro-Kac-Moody primary satisfies $h_r \le h_1$ we reach a contradiction, as any other state satisfies $h_r \le h_1 < h$ and gives a positive contribution to the crossing equation. Thus, the weight of the first excited primary state is bounded by $h_1 < h_r$. This result holds provided the spectrum is bounded from below. Otherwise, states with arbitrarily low conformal weight exist for which $G(h,\k)$ is negative and the crossing equation is satisfied (see Fig.~\ref{gfunction}). 

The value of $h_r$ depends on the characters $\chi_{h_0,p_0}(\tau,z)$ and $\chi_{h_1,p_1}(\tau,z)$. Using the identities and special values of the Dedekind Eta function given in Appendix~\ref{ap:modular} we find that the (weakest) bound on the first excited state is given by
  \eq{
  h_1 < \frac{c}{12} + 0.479, \qquad \qquad c \gg 1.\label{hellerman2}
  }
This bound corresponds to the case where the characters of both the $h_1$ and $h_0$ states are given by eq.~\eqref{realcharacter} and $h_0$ is assumed to be the vacuum. Hence Virasoro-Kac-Moody primaries with negative norm descendant states yield the weakest bound at subleading order in $c$. On the other hand, when the characters of these states are given by eq.~\eqref{imaginarycharacter}, i.e.~when the charges $p_1$ and $p_0$ are assumed to be imaginary, and neither of them is the vacuum, we obtain a stronger bound $h_1 < c/12 + (12-\pi)/6\pi \approx c/12 + 0.470$. For any other combination of characters the corresponding bound on the weight $h_1$ lies between this value and that given in eq.~\eqref{hellerman2}. Thus, the presence of negative norm descendant states weakens the bound at most by terms of $\O(10^{-3})$. 

We conclude this section by noting that eq.~\eqref{hellerman2} applies to any CFT with a $\U1$ algebra featuring a negative level, positive norm Virasoro-Kac-Moody primaries, and the constraints on the conformal weight given in eq.~\eqref{unitaritybound}. The positivity of the $D_{h,p}$ coefficients is an important assumption as it led to the addition of states with imaginary charge. If one demands reality of all charges, then at least one Virasoro-Kac-Moody primary must have negative norm and the bound derived above may still hold but can no longer be justified.


\section{Unitary WCFT bounds} \label{se:morebounds}

While our focus has been on holographic WCFTs, we will not shy away from deriving bounds on unitary WCFTs with a positive level. The analysis is facilitated by the fact that the characters of unitary Virasoro-Kac-Moody representations with positive central charge, positive level, and $h \ge h_0 = 0$ are independent of the charge of the state. Indeed, the unitary Virasoro-Kac-Moody character is already given in eq.~\eqref{imaginarycharacter} and reproduced here for convenience
  \eq{
  \chi_{h, p} (\tau,z) = \frac{1}{\eta(\tau)^2}q^{h - (c-2)/24} y^p (1 - \d_{vac} q). \label{unitarycharacter}
  }
Now note that the crossing equation~\eqref{crossing5} used in the previous section holds for theories with a positive or negative level, as its dependence on the $\u1$ level drops out at the fixed point $\tau = i$, $z = 0$. Furthermore, the corresponding $G(h,\k)$ function is still given by eq.~\eqref{G} where the omitted terms can be computed using the results of Appendix~\ref{ap:modular}. Thus, assuming only Virasoro-Kac-Moody primaries with positive norm are featured in the spectrum, we find that the first excited primary state satisfies
  \eq{
  h_1  < \frac{c}{12} + 0.471, \qquad \qquad c \gg 1.
  }
Not surprisingly, this bound is stronger from that found in holographic WCFTs by a term of $\O(10^{-3})$, a consequence of the subleading nature of descendant states which distinguish different characters.

Let us now adapt the bounds derived in refs.~\cite{Benjamin:2016fhe,Dyer:2017rul} to unitary WCFTs with a positive level. We first note that the modular S-transformation of the WCFT partition function given in eq.~\eqref{spartition} is indistinguishable from that of a CFT with an internal $\u1$ symmetry~\cite{Detournay:2012pc,Kraus:2006nb}. However, for WCFTs with a positive level we must flip the sign of the exponential term in eq.~\eqref{spartition}. We now show that, for $\tau = i \t$ and up to a factor of $(1 - \d_{vac} e^{-2\pi \t})$, the WCFT partition function~\eqref{partitioncharacters} is equivalent to the CFT partition function considered in ref.~\cite{Dyer:2017rul}. The equivalence is established after an appropriate identification of their parameters (on the left of the equations below)
  \eq{
  \tilde{\dd} \ra h, \qquad \qquad \tilde{c} \ra \frac{c}{2}. \label{partitionmap}
  }
The reason the partition functions match up to a factor of $(1 - \d_{vac} e^{-2\pi \t})$ is that ref.~\cite{Dyer:2017rul} considers a representation of the partition function in terms of Virasoro but not Kac-Moody characters. That is, their partition function $\tilde{Z}(\tau,\bar{\tau},z)$ at $\tau = i \t$ reads
  \eq{
  \tilde{Z}(\tau,\bar{\tau},z) &= \sum_{\tilde{h}, \tilde{\bar{h}},p} D_{\tilde{h},\tilde{\bar{h}},p} \frac{q^{\tilde{h} - (\tilde{c} - 1)/24}y^p }{\eta(\tau)}  (1- \d_{vac} q)\frac{
\bar{q}^{\tilde{\bar{h}} - (\tilde{c} - 1)/24}}{\eta(\tau)^*}  (1- \d_{vac} \bar{q}),  \label{dyer0} \\
& = \sum_{h,p} D_{\tilde{h},\tilde{\bar{h}},p}  \frac{1}{\eta(\tau)^2} q^{ h - (c - 2)/24}y^p  (1- \d_{vac} q)^2 \Big |_{\tau = i\t}, \label{dyer1}
  } 
where $\bar{q} = e^{-2\pi i\bar{\tau}}$ and we have used eq.~\eqref{partitionmap} with $\tilde{\Delta} = \tilde{h} + \tilde{\bar{h}}$. Clearly, up to the additional factor of $(1 - \d_{vac} e^{-2\pi \t})$ accompanying the vacuum CFT character, eq.~\eqref{dyer1} shares the same structure as the partition function of a unitary WCFT with character~\eqref{unitarycharacter}. In particular, note that although $D_{\tilde{h},\tilde{\bar{h}},p}$ in eq.~\eqref{dyer1} and $D_{h,p}$ in the WCFT partition function count different kinds of states, all that is required by the modular bootstrap is that these coefficients are positive semidefinite.
  
The presence of an extra $(1 - \d_{vac} e^{-2\pi \t})$ term is not discouraging, for it has a similar effect as that of the negative norm states considered in Section~\ref{se:virasorobound}. Furthermore, note that ref.~\cite{Benjamin:2016fhe} considers a representation of the partition function in terms of primary and descendant states, i.e.~not in terms of characters. In the Hellerman bound for holographic WCFTs~\eqref{hellerman2} different characters lead to $\O(10^{-3})$ corrections, the latter of which are small due to the small number of $\tau$ derivatives taken in the crossing equation. For this reason any discrepancy between characters should not affect the bounds reported in refs.~\cite{Benjamin:2016fhe,Dyer:2017rul} when the number of $\tau$ derivatives is small. Thus, in the large central charge limit the spectrum of a modular invariant, unitary WCFT must feature~\cite{Benjamin:2016fhe,Dyer:2017rul}
  \eq{
\frac{h}{|p|} &\le \sqrt{2\pi} \frac{c}{6} + \O(1), && \textrm{for at least one state,}  \\
  h & < \frac{c}{12} + \O(1), \qquad&& \textrm{for the lightest charged state,} \\
 |p| &\le \frac{1}{\sqrt{2}}, && \textrm{for the smallest possible charge,}
  }
where the relative factors of $\sqrt{2}$ with respect to refs.~\cite{Benjamin:2016fhe,Dyer:2017rul} follow from our conventions for the Virasoro-Kac-Moody algebra. 

We conclude this section by noting that ref.~\cite{Dyer:2017rul} provides numerical evidence for a stronger bound on the conformal weight of the lightest charged state. This bound is derived by extrapolating to infinity the number $N$ of $\tau$ derivatives of the crossing equation obtained from eq.~\eqref{dyer0}. As shown in Table~1 of ref.~\cite{Dyer:2017rul}, increasing the number of derivatives at fixed central charge gives only subleading corrections to the bound, a fact that makes the extrapolation $N \ra \infty$ possible. In particular, Table~1 of ref.~\cite{Dyer:2017rul} suggests that at large but fixed $c$, the bound on the lightest charged state can be derived using only up to $N_c$ derivatives where ${N_c/c}$ is small. In our case the difference between vacuum characters in the WCFT and CFT partition functions grows with the number of $\tau$ derivatives taken on the characters. However, up to $N_c$ derivatives, we still expect the relative factor of $(1 - \d_{vac} e^{-2\pi \t})$ to yield subleading contributions to the bound, as in Table~1 of ref.~\cite{Dyer:2017rul}. Thus, unitary WCFTs should satisfy a stronger bound on the lightest charged state in the large central charge limit, namely~\cite{Dyer:2017rul}
  \eq{
  h & < \frac{c}{\a} + \O(1), \,\, \qquad \qquad  \a > 16.
  }
  %


\section{Bulk interpretation of holographic WCFT bounds} \label{se:bulk}

We now turn to the holographic interpretation of the results derived in Section~\ref{se:bounds}, beginning with the constraints on the $\u1$ charges.


\subsubsection*{Imaginary \texorpdfstring{$\u1$}{U(1)} charges}

\begin{figure}[!h]
  \centering
  \includegraphics{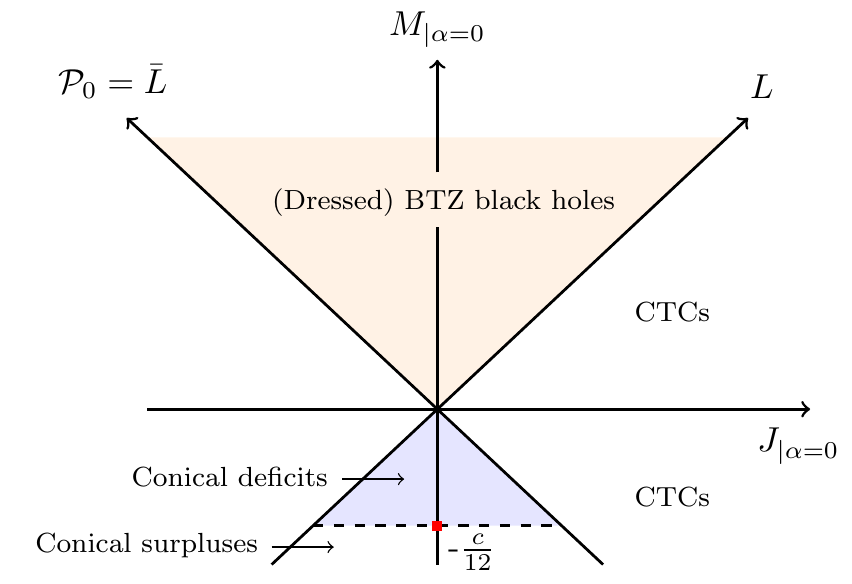}
  \caption{With the exception of the vacuum, states with imaginary $\u1$ charge in the WCFT correspond to pathological solutions with $\bar{L} < 0$ in the bulk theory.}
  \label{spectrum2}
  \end{figure}

The states with imaginary charge required by the modular bootstrap have an interesting bulk interpretation. This follows from the relationship between zero modes of the bulk and boundary algebras~\eqref{zeromodesmap} whereby states with imaginary charge correspond to bulk metrics with $\bar{L} < 0$. One such state is the AdS$_3$ vacuum which, despite not saturating the bound on the conformal weight given in eq.~\eqref{unitaritybound}, is annihilated by the generators of global symmetries. The results of Section~\ref{se:u1bounds} suggest that if the quantum theory of 3D gravity with CSS boundary conditions contains Virasoro-Kac-Moody primaries with positive norm, then it must also include at least one solution other than the vacuum and (healthy) BTZ black holes. With the exception of the AdS$_3$ vacuum, the solutions with $\bar{L} < 0$ include metrics with either (see Fig.~\ref{spectrum2})
\begin{itemize}
\item[\emph{i.}] \emph{causal singularities (CTCs)} if $0 < L$~\cite{Brown:1986nw,Banados:1992gq}, 
\item[\emph{ii.}] \emph{conical deficits} if $L \le 0$ and $L + \bar{L} > - c/12$~\cite{Deser:1983nh},
\item[\emph{iii.}] \emph{and conical surpluses} if $L \le 0$ and $L + \bar{L} < - c/12$~\cite{Deser:1983nh}.
\end{itemize}
Conical defects stand out among these solutions as they result from the presence of matter in the bulk theory.

Our results are reminiscent of the conclusions reached in ref.~\cite{Maloney:2007ud} which show that 3D gravity with Brown-Henneaux boundary conditions but without matter, and in fact without any of the pathological $\bar{L} < 0$ solutions, is inconsistent quantum mechanically. While both our analysis and that of ref.~\cite{Maloney:2007ud} exploit modular invariance of the partition function, they do so in different ways. In this paper we have assumed that the $D_{h,p}$ coefficients of the partition function are positive integers and used modular invariance to obtain an inconsistent theory unless states with imaginary charge are added to the spectrum. In contrast, ref.~\cite{Maloney:2007ud} assumes a spectrum for 3D gravity already compatible with modular invariance, consisting of the vacuum and all of its modular images, and obtains $D_{h,\bar{h}}$ coefficients that do not admit a physical interpretation.

Thus, a possible interpretation of our results is that 3D quantum gravity with a negative cosmological constant must feature additional solutions beyond the vacuum and BTZ black holes (including their $SL(2,\mathbb{Z})$ cousins~\cite{Maldacena:1998bw,Maloney:2007ud}), regardless of whether one imposes Brown-Henneaux or CSS boundary conditions.


\subsubsection*{Bounds on angular momentum}

The bound on the conformal weights imposed in eq.~\eqref{unitaritybound} has dramatic consequences in the bulk theory. Indeed, using the map between zero modes given in eq.~\eqref{zeromodesmap}, we obtain the following constraints on the bulk data~\eqref{metric},
  \eqsp{
  L &\ge -\frac{c}{24}, \qquad \quad\textrm{if } \bar{L} < 0, \\
  L - \bar{L} & \ge - \frac{c}{24}, \qquad \quad \textrm{if } \bar{L} \ge 0. \label{unitarityboundbulk}
  }
The first constraint is also found in 3D gravity with Brown-Henneaux boundary conditions while the second constraint is more novel, although not completely unexpected given that the Sugawara contribution to the $L_0$ mode in eq.~\eqref{Lncanonical} has an overall negative sign. 

Note that eq.~\eqref{unitaritybound} places no restriction on $B'(x^+)$, one of the components of the boundary metric, which is left arbitrary. The reason for this is that metrics with nonvanishing $B'(x^+)$ correspond to nonlinear solutions of 3D gravity dressed with $\U1$ descendants. This observation follows directly from the linear dependence of the bulk $\P_n$ charges on $B'(x^+)$. In contrast, eq.~\eqref{unitaritybound} constrains only the weight of Virasoro-Kac-Moody primaries, and has little to say about the corresponding descendant states. Hence eq.~\eqref{unitaritybound} can only restrict the charges of the $B'(x^+) = 0$ solutions of the bulk theory. In that case the parameter $\a^2$ introduced in eq.~\eqref{MJ} vanishes and we obtain the following constraints on the charges
  \eqsp{
  \L_0{} _{|\a = 0} &\ge -\frac{c}{24}, \qquad \quad \textrm{if } \bar{L} < 0,\\
  J _{|\a=0} &\ge -\frac{c}{24}, \qquad \quad \textrm{if } \bar{L} \ge 0. \label{Jbound}
  }

\begin{figure}[!h]
   \begin{center}
   \includegraphics{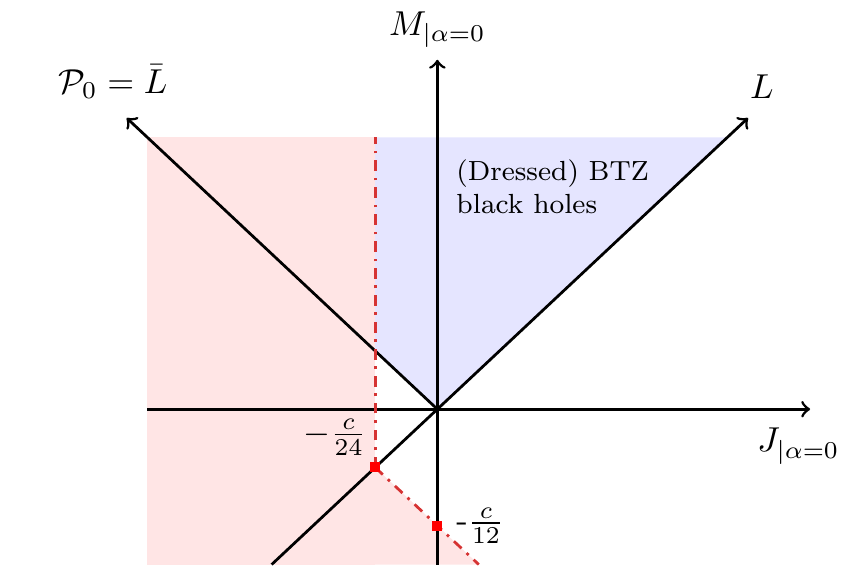}
   \end{center}
   \caption{The bound~\eqref{unitaritybound} restricts the parameters $L$ and $\bar{L}$ of the bulk theory to lie outside the red region (left of the dash-dotted line). When $B(x^+) = 0$, eq.~\eqref{unitaritybound} puts a lower bound on the angular momentum of BTZ black holes, i.e.~$J_{|\a = 0} \ge -\frac{c}{24}$.}
   \label{unitaryspectrum}
   \end{figure}

The second constraint in eq.~\eqref{Jbound} puts a lower bound on the angular momentum of BTZ black holes, among other undressed solutions, as illustrated in Fig.~\ref{unitaryspectrum}. From the WCFT point of view, the solutions with $\bar{L} > 0$ and angular momentum $J _{|\a=0} < -c/24$ must feature negative norm Virasoro descendants. These states are interpreted as boundary gravitons in the bulk theory that are crucially different from the standard boundary gravitons of Brown and Henneaux, due to the CSS boundary conditions. It would be interesting to understand the physical implications of including such states in the spectrum of 3D gravity and a few speculations are considered at the end of this section. Relatedly, note that the CSS boundary conditions break chirality by promoting only the $\g_{++} = B'(x^+)$ component of the boundary metric to a dynamical variable. The ``conjugate'' choice, namely $\g_{--} = C'(x^-)$, exchanges the $L$ and $\bar{L}$ dependence of the bulk charges and, in particular, that of eq.~\eqref{unitarityboundbulk}. This alternative choice of CSS boundary conditions leads to a complimentary upper bound on the angular momentum of the BTZ black holes which must then satisfy $J_{|\a = 0} \le c/24$.

The Hellerman bound~\eqref{hellerman2} has an equally interesting but by now not surprising bulk interpretation. Using the map between the bulk and boundary zero modes~\eqref{zeromodesmap} the angular momentum of the first excited primary state besides the $\u1$ current is bounded from above by
  \eq{
    J _{|\a=0} =  L - \bar{L} < \frac{c}{24} + 0.479. \label{hellermanbulk}
  }
Note that eq.~\eqref{hellermanbulk} is a direct consequence of the bound imposed on the conformal weight in eq.~\eqref{unitaritybound}. The latter guarantees a bounded spectrum that leads to a well defined and, from the point of view of the WCFT, physically motivated partition function. If the bound~\eqref{unitaritybound} is not saturated we find instead
  \eq{
  J _{|\a=0} < \frac{c}{24} - h_0 + 0.479,
  }
for some $h_0 < c/24$. In any case, the curious map between the bulk and boundary algebras implies that any bounds on the conformal weight of the WCFT correspond to bounds on the angular momentum of undressed solutions in the bulk theory.

\subsubsection*{Comments on boundedness of the spectrum}

We have seen that a bounded spectrum in the WCFT has dramatic consequences
in the dual theory of gravity. Note, however, that it is possible to relax eq.~\eqref{unitaritybound} and extend the $h \ge -p^2$ bound to all states. In this case we should consider the algebra that results from subtraction of the Sugawara stress tensor, which is characterized by a Virasoro zero mode $L^{(s)}_0$ that is bounded from below, cf.~eq.~\eqref{sugawarabasis}. We can then define a new partition function where it is possible to derive a new Hellerman bound for the eigenvalue of $L^{(s)}_0$. In the bulk this bound implies that the first excited state satisfies instead
  \eq{
  L = M + J < \frac{c}{24} + \O(1),
  }
which is the same bound one obtains from a CFT dual to three-dimensional gravity with Brown-Henneaux boundary conditions.

Nevertheless, it would be interesting to understand how the lower bound on the angular momentum given in eq.~\eqref{Jbound} may arise in the gravitational context. One obvious possibility is that the bound is a technical requirement, one that is necessary to validate the Hellerman bound derived in the WCFT via the modular bootstrap. Alternatively, the bound~\eqref{Jbound} may be an artifact of the nonlocal map between the CSS and canonical Virasoro-Kac-Moody algebras which should be understood better. Another possibility is that~\eqref{Jbound} is required for consistency of the bulk theory, the latter of which already violates chirality through the boundary conditions. Finally, it is possible that backgrounds with $J _{|\a=0} < -c/24$ and $\bar{L} > 0$ feature instabilities triggered by the addition of descendant states. If so it would be interesting to find the mechanism responsible for such instabilities, which would also explain the origin of the negative level in the Virasoro-Kac-Moody algebra.


\section{Conclusions} \label{se:discussion}  

In this paper we exploited modular covariance of the partition function to derive constraints on the spectrum of warped CFTs. We found that despite the mild violation of unitarity, the modular bootstrap is still feasible in theories with negative norm descendant states. This allowed us to constrain the spectrum of holographic WCFTs and interpret our results in the dual theory of gravity. We have argued that these results are universal and apply to any theory with a Virasoro-Kac-Moody algebra and negative level satisfying natural assumptions on boundedness and the norm of primary states. Furthermore, the fact that conventional CFTs with internal $\u1$ symmetries share the same modular transformation properties as WCFTs allowed us to extend the bounds derived in~\cite{Benjamin:2016fhe,Dyer:2017rul} to unitary WCFTs with a positive level.

There are in principle no obstructions to the application of more advanced numerical methods to further constrain the spectrum of holographic WCFTs. In particular, we expect that higher order differential operators can be used to prove the existence of additional states with imaginary charge. Furthermore, we expect to find a similar improvement to the bound on the first excited state reported in refs.~\cite{Collier:2016cls,Dyer:2017rul}. Likewise, it should be possible to derive bounds on the lightest charged state of holographic WCFTs, as well as bounds on the weight-to-charge ratio of at least one state. One complication in this analysis is the existence of states with imaginary charge since the crossing equation receives contributions of $\O(p^2)$ whose sign depends on the spectrum, i.e.~on whether the charge is real or imaginary. This requires the construction of higher order differential operators $\dd$ featuring both $z$ and $\tau$ derivatives.



\section*{Acknowledgments}
We are grateful to Pankaj Chaturvedi, Bin Chen, Pengxiang Hao, Diego Hofman, and Jianfei Xu for helpful discussions. In particular, we are thankful to Diego Hofman for comments on the draft. The authors thank the Tsinghua Sanya International Mathematics Forum for hospitality during the workshop and research-in-team program ``Black holes, Quantum Chaos, and Solvable Quantum Systems''. This work was supported by the National Thousand-Young-Talents Program of China and NFSC Grant No.~11735001. The work of L. \!A. was also supported by the International Postdoc Program at Tsinghua University. 


\appendix

\section{State-dependent asymptotic Killing vectors} \label{ap:dhh}

In this appendix we show that the nonlocal map between the CSS~\eqref{cssu1} and canonical~\eqref{canonicalu1} Virasoro-Kac-Moody algebras can be derived using the state-dependent asymptotic Killing vectors found in~\cite{Compere:2013bya}. The corresponding charges are well-defined, i.e.~finite, conserved, and integrable, provided that the function $B(x^+)$ in eqs.~\eqref{boundarymetric} and~\eqref{metric} is periodic. Crucially, with this choice of asymptotic Killing vectors the zero mode charge $\P_0$ is allowed to vary, which is an important assumption made in the analysis of the dual field theory. 

The $\L_n$ and $\P_n$ charges given in eqs.~\eqref{Ln} and~\eqref{Pn} correspond to the symmetries generated, respectively, by the following asymptotic Killing vectors~\cite{Compere:2013bya}
   \eq{
   \xi_n = e^{i n x^+} \( \p_+ - \frac{r}{2} i n \p_r  \) + \dots , \qquad \qquad \eta_n = e^{i n x^+} \p_- + \dots ,
   }
where $\dots$ denote terms subleading in the radial coordinate $r$. The derivation of the charges in eqs.~\eqref{Ln} and~\eqref{Pn} assumes that the $\P_0 = \bar{L}$ charge is a fixed constant. To remedy this, the following state-dependent asymptotic Killing vectors were proposed in Appendix B of ref.~\cite{Compere:2013bya}\footnote{State-dependent Killing vectors had been discussed earlier in the context of warped AdS$_3$ in ref.~\cite{Compere:2008cv}.} 
  \eq{
  \xi_n \ra \xi'_n = \xi_n - \eta_n, \qquad \qquad \eta_n \ra \eta_n' = \frac{\eta_n}{2\sqrt{\bar{L}}}, \label{killing2}
  }
where we have rescaled $\eta'_n$ by a factor of 2 with respect to~\cite{Compere:2013bya}. Let us denote the charges corresponding to $\xi'_n$ and $\eta'_n$ by $L_n$ and $P_n$, respectively. The latter are given by
  \eq{
  L_n & = \frac{1}{2\pi} \int d\phi \, e^{inx^+} \Big \{ L(x^+) - \bar{L} \big[ 1 + B'(x^+) \big]^2  \Big \},  \label{apLn} \\
  P_n & = \frac{1}{2\pi} \int d\phi \, e^{inx^+}  \Big \{ \sqrt{\bar{L}} + \sqrt{\bar{L}} B'(x^+) \Big \}, \label{apPn}
  }
and satisfy the Virasoro-Kac-Moody algebra~\eqref{canonicalu1} with central charge $c = 3\ell/2G$ and level $k = -1$. Both the $L_n$ and $P_n$ charges are integrable for varying $\bar{L}$. Furthermore, note that the second term in eq.~\eqref{apLn} is precisely the Sugawara contribution of the $P_n$ charges, as noted in eq.~\eqref{Lncanonical} for the $L_n$ and $P_n$ charges defined in eq.~\eqref{mapu1s}. In fact, using eqs.~\eqref{Ln} and~\eqref{Pn}, it is not difficult to show that the $L_n$ and $P_n$ charges given in eqs.~\eqref{apLn} and~\eqref{apPn} satisfy
  \eq{
  \L_n = L_n + 2 P_0 P_n - P_0^2 \,\d_n, \qquad \qquad \P_n = 2 P_0 P_n - P_0^2 \,\d_n,\label{apmapu1s}
  }
which is the nonlocal map between the Virasoro-Kac-Moody charges given in eq.~\eqref{mapu1s}. In particular, note that due to the square root in the definition of $P_n$ in eq.~\eqref{apPn}, the charges of states with $\bar{L} < 0$ are antihermitian and satisfy eq.~\eqref{antihermitianP}.

Thus, by means of the state-dependent asymptotic Killing vectors~\eqref{killing2} we can kill two birds with one stone. On the one hand we render the charges of 3D gravity with CSS boundary conditions finite, conserved, and integrable for varying $\bar{L}$. On the other hand we reproduce the nonlocal map of ref.~\cite{Detournay:2012pc} that is used to put the algebra in eq.~\eqref{cssu1} into its canonical form~\eqref{canonicalu1}.


\section{Useful formulae} \label{ap:modular}

In this appendix we collect useful formulae necessary to derive the Hellerman bound. In principle it is possible to simplify the crossing equation by multiplying the partition function by appropriate powers of $\eta(\tau)$, as done in ref.~\cite{Collier:2016cls}. However, since the Virasoro-Kac-Moody character is different for real and imaginary charges, it is more convenient to leave intact all instances of $\eta(\tau)$ appearing in the partition function. 

Now note that derivatives of the Dedekind eta function are linear in $\eta(\tau)$ and proportional to powers of the Eisenstein series $E_n(\tau)$ where $n = 2, 4, 6$. This follows from 
  \eq{
  \p_{\tau} \eta(\tau) &= \frac{i\pi}{12} \eta(\tau) E_2(\tau),
  }
and the Ramanujan identities,
  \eq{
  \p_{\tau} E_2(\tau) &= \frac{\pi i}{6} \big [ E_2(\tau)^2 - E_4(\tau) \big ], \\
  \p_{\tau} E_4(\tau) &= \frac{2\pi i}{3} \big [ E_2(\tau) E_4 (\tau) - E_6(\tau) \big],\\
  \p_{\tau} E_6(\tau) &= \pi i \big [ E_2(\tau) E_6 (\tau) - E_4(\tau)^2 \big].
  }
Next, we note that at the fixed point $\tau = i$ the values of the Dedekind eta function and the Eisenstein series can be shown to satisfy
  \eq{
  \eta(i) = \frac{\g}{2 \pi^{3/4}}, \qquad E_2(i) = \frac{3}{\pi}, \qquad E_4(i) = \frac{3}{(2\pi)^6}  \g^8, \qquad E_6(i) = 0, \label{identitiesati}
  }
where $\g$ is given by
  \eq{
  \g = \GG\(\tfrac{1}{4}\).
  }
Since the character of Virasoro-Kac-Moody representations with negative level and real charge is inversely proportional to $\eta(2\tau)$, it is also necessary to know that 
  \eqsp{
  \eta(2i) &= \frac{\g}{2^{11/8} \pi^{3/4}},
  }
and the corresponding values of the Eisenstein series which are given by
  \eq{
  \qquad E_2(2i) = \frac{3}{2^{5}\pi^3} (2^4\pi^2 + \g^4), \qquad E_4(2i) = \frac{33}{2^{10} \pi^6} \g^8 , \qquad E_6(2i) &= \frac{189}{2^{15}\pi^9}  \g^{12}. \label{identitiesat2i}
  }
The values of $\eta(\tau)$ in eqs.~\eqref{identitiesati} and~\eqref{identitiesat2i} can be found in the fifth volume of Ramanujan's notebooks~\cite{Berndt:1998ab}. On the other hand, the values of the Eisenstein series may be obtained from the observation that any modular form, including $E_2(\tau) - 2 E_2(2\tau)$ which is modular at level 2, can be written in terms of quotients of $\eta(\tau)$, $\eta(2\tau)$, and $\eta(4\tau)$~\cite{Ono:2004ab,Kilford:2007ok}.\footnote{Note that ref.~\cite{Kilford:2007ok} contains a typo and $E_2(\tau) - 2 E_2(2\tau) = - \eta (2 \tau )^{20} / [\eta (\tau ) \eta (4 \tau )]^8 - 16 \eta (4 \tau )^8 /  \eta (2 \tau )^4$.}
  


\ifprstyle
	\bibliographystyle{apsrev4-1}
\else
	\bibliographystyle{utphys2}
\fi

\bibliography{warpedbound}



\end{document}
